\documentclass[%
 reprint,
 superscriptaddress,
 amsmath,amssymb,
 aps,
 pra
]{revtex4-2}

\usepackage{graphicx}
\usepackage{dcolumn}
\usepackage{bm}
\usepackage{hyperref}

\begin{document}

\title{Thermodynamics in expanding shell-shaped Bose-Einstein condensates}

\author{Brendan Rhyno}
\email{brhyno2@illinois.edu}
\affiliation{
Department of Physics, University of Illinois at Urbana-Champaign, Urbana, Illinois 61801-3080, USA
}

\author{Nathan Lundblad}
\affiliation{
Department of Physics and Astronomy, Bates College, Lewiston, ME, 04240, USA
}

\author{David C. Aveline}
\affiliation{
Jet Propulsion Laboratory, California Institute of Technology, Pasadena, CA, 91109, USA
}

\author{Courtney Lannert}
\email{clannert@smith.edu}
\affiliation{
Department of Physics, Smith College, Northampton, Massachusetts 01063, USA
}
\affiliation{
Department of Physics, University of Massachusetts, Amherst, Massachusetts 01003-9300, USA
}

\author{Smitha Vishveshwara}
\email{smivish@illinois.edu}
\affiliation{
Department of Physics, University of Illinois at Urbana-Champaign, Urbana, Illinois 61801-3080, USA
}

\date{\today}

\begin{abstract}

Inspired by investigations of Bose-Einstein condensates (BECs) produced in the Cold Atom Laboratory (CAL) aboard the International Space Station, we present a study of thermodynamic properties of shell-shaped BECs.  Within the context of a spherically symmetric `bubble trap' potential, we study the evolution of the system from small filled spheres to hollow, large, thin shells via the tuning of trap parameters. We analyze the bubble trap spectrum and states,  and track the distinct changes in spectra between radial and angular modes across the evolution. This separation of the excitation spectrum provides a basis for quantifying dimensional cross-over to quasi-2D physics at a given temperature. Using the spectral data, for a range of trap parameters, we compute the critical temperature for a fixed number of particles to form a BEC. For a set of initial temperatures, we also evaluate the change in temperature that would occur in adiabatic expansion from small filled sphere to large thin shell were the trap to be dynamically tuned. We show that the system cools during this expansion but that the decrease in critical temperature occurs more rapidly, thus resulting in depletion of any initial condensate. We contrast our spectral methods with standard semiclassical treatments, which we find must be used with caution in the thin-shell limit. With regards to interactions, using energetic considerations and corroborated through Bogoliubov treatments, we demonstrate that they would be less important for thin shells due to reduced density but vortex physics would become more predominant. Finally, we apply our treatments to traps that realistically model CAL experiments and borrow from the thermodynamic insights found in the idealized bubble case during adiabatic expansion.

\end{abstract}

\maketitle

\section{Introduction}

The physics of Bose-Einstein condensates that form closed shell-like geometries is a fascinating and far-reaching topic which spans a range of physical scales and phenomena.
On astronomic scales, for instance,  portions of neutron star interiors could potentially contain superfluid spherical-shell structures \cite{Sauls_1989,Pethick_2017}.
Exotic stars known as ``boson stars" have also been hypothesized to form when a complex scalar field couples to gravity \cite{Liebling_2017}.
On the microscopic length scales of trapped ultracold atoms, concentric shells of differing phases can be generated in the setting of Bose-Fermi mixtures \cite{Molmer_1998,Ospelkaus_2006,Schaeybroeck_2009,Lous_2018} as well as in bosonic optical lattices, which can preferentially favor superfluid versus Mott insulating phases in different regions \cite{Greiner_2002,Batrouni_2002,DeMarco_2005,Campbell_2006,Barankov_2007,Sun_2009}.
The prospect of realizing isolated shell-shaped condensates has received a surge of interest \cite{Zobay_2001,Zobay_2004,Colombe_2004,DeMarco_2006,Merloti_2013,Garraway_2016,Perrin_2017,Harte_2018,Aveline_2018,Meister_2019,Lundblad_2019,Condon_2019,Aveline_2020,Frye_2021}, as has investigations into the nature of the onset of condensation in other novel geometric settings \cite{van_Druten_1997,Chomaz_2015}.
However, on Earth, gravity renders such a realization challenging by causing trapped gases to pool at the bottom of the trap. Terrestrial experiments done in free-fall can mitigate this problem \cite{Zoest_2010,Muntinga_2013,Condon_2019}, but the inherently short condensate lifetimes are undesirable. It would thus be ideal to probe shell-shaped condensates in perpetual free-fall. The International Space Station (ISS) provides precisely these conditions, operating in a microgravity environment.

Bose-Einstein condensate (BEC) shells in isolation would provide an arena for studying numerous interesting features which could translate to salient properties of shells in these various settings. The topology of hollowed-out fluid structures shows innate differences from fully-filled structures. The presence of an inner and an outer boundary affects collective mode spectra, vortex physics, and thermodynamics. With regards to geometry, all these features show unique characteristic properties as a spherical system undergoes an evolution from filled to slightly hollowed out to the thin-shell limit, including tell-tale signatures of topological change. In principle, this evolution can be realized by forming a condensate in a trap that can produce the standard filled geometry, and then, as exemplified by the `bubble trap' \cite{Zobay_2001,Zobay_2004}, the hollowing out can take place by tuning trap parameters. A major aspect of BEC shells, which we address here, is the thermodynamics behind how a hollow shell condensate structure can be created at finite temperature. Towards actual realization of such structures, our theoretical analyses closely target the experiments being conducted by two of us (D. A. and N. L.) aboard the ISS. 

In 2018 the Cold Atom Laboratory (CAL), developed by the Jet Propulsion Laboratory was successfully launched into orbit aboard the ISS. CAL's design allows for remote generation of BECs in microgravity \cite{Aveline_2018,Aveline_2020} and has been able to produce large millimeter scale ultracold bubbles in the $10-100~\text{nK}$ temperature range \cite{Lundblad_2021}. CAL offers wide capabilities for generating ultracold mixtures \cite{Aveline_2018}, but important to this work is its ability to generate so-called ``dressed potentials" \cite{Lundblad_2019}. In this mode of operation, an initial non-hollow condensate is prepared using magnetically trapped $^{87}\text{Rb}$ atoms in the $|F=2,m_F=2\rangle$ hyperfine state \cite{Lundblad_2019}. A radio frequency (rf) signal is then turned on which modifies the potential experienced by the atoms. By modifying this rf ``detuning" signal, one can produce potentials capable of harboring BEC shells.
The first set of experiments reveal a dramatic expansion and hollowing-out of an initial condensate as it undergoes significant thermodynamic changes \cite{Lundblad_2021}.

In this work, complementing the CAL experiments, we theoretically investigate the thermodynamic process of a trapped Bose gas, initially forming a compact filled sphere, undergoing adiabatic expansion due to changing trap parameters to form a large, thin spherical shell. Specifically, we consider two aspects of this process -- in a bubble trap geometry, for fixed number of gas particles, how does the temperature of the system evolve under adiabatic expansion? How does this contrast with the critical temperature for condensation for a given set of trap parameters? We also perform an analysis for the experimental traps employed on CAL. Recent work has begun to address related issues; aspects of a thin spherical shell's thermodynamics have been investigated \cite{Tononi_2020,Greghi_2020}, including the fate of the Berezinskii-Kosterlitz-Thouless transition in a closed, thin-shell geometry \cite{Tononi_2021}.

Our findings here make for a comprehensive thermodynamic study of this unique structure and evolution in and of itself while also informing the related experiments. They predict the manner in which the temperature of the Bose-gas would change during the adiabatic hollowing-out expansion. Comparing with the local critical temperature throughout the process, we are led to conclude that there is a narrow window of parameters in which a large thin condensate shell can be created. We go beyond the semiclassical approximation,
whose predictions should be taken with care in quasi-2D systems \cite{van_Druten_1997,Chomaz_2015}.
We also include, at first order, the effect of interactions and show that they are largely unimportant in the same large thin limit. Our results, while predicting that the first set of shell-potential experiments aboard the ISS are unlikely to retain condensation at the larger radii, point to the range of parameters wherein large condensate shells occur. They also ascertain the stability of the spectacular systems that these first experiments exhibit: compared to regular trapped gases of micron-scale, thousand-fold adiabatic expansion gives rise to exquisite delicate gigantic thermal gas bubbles.

This paper is organized as follows: we first consider bosons subject to a radial ``bubble trap" potential \cite{Zobay_2001,Zobay_2004}. After exploring the spectral properties of bubble-trapped states, we use this information to calculate thermodynamic quantities in the noninteracting case. In particular, we determine the BEC critical temperature as the potential evolves from a filled-to-hollow sphere geometry along with the temperature of the system when the expansion process is performed adiabatically. Strikingly, we find that adiabatic expansion of a BEC leads to condensate depletion. We then contrast our results with the semiclassical approximation and find semiclassical predictions overestimate the critical temperature and hence partially conceal the condensate depletion phenomena.
Next, we consider the effects of interactions and dimensionality. Contrasting zero-temperature Gross-Pitaevskii numerics with various approximation schemes reveals that, for fixed particle number systems, the decreasing density of an expanding bubble means the noninteracting description is well-suited to describe thin-shell geometries.
We then use this result to construct a Bogoliubov description of collective excitations in this instance. Consistent with the above discussion, we find thermodynamic predictions for thin shells are relatively close to those in the absence of interactions.
By construction, the Bogoliubov effective theory assumes excitations dominated by long-wavelength phase and density fluctuations above a $U(1)$ symmetry broken ground state; we indicate the eventual breakdown of this description as one crosses over into the quasi-2D limit.
Finally, we apply the lessons of the bubble trap to compute thermodynamics in trapping potentials achieved with CAL.
We conclude with an outlook on how our work can inform the CAL experiments and discuss possible manifestations of non-equilibrium physics.

\section{Bubble trap spectrum and states}

Here we introduce the bubble trap as means to continuously tune the system geometry. We then discuss the properties of states subject to the bubble trap potential as it deforms from a filled sphere to a hollow thin shell.

\subsection{Trapping potential}

Our focus is the thermodynamics of ultracold bosonic gases trapped in potentials that allow geometries ranging from a filled sphere to a nearly two-dimensional hollow shell. Such dimensional crossovers have been investigated employing hard-wall spherical potentials and confinement to spherical surfaces \cite{Bereta_2019,Tononi_2019}. Here we employ trapping potential forms that can be continuously tuned to span the whole range and can approximate the related experimental setting of shell-shaped BECs aboard CAL \cite{Lundblad_2019}.

As our starting point, we model the dilute collection of trapped, interacting bosons in three dimensions using the standard description provided by the Hamiltonian:
\begin{eqnarray}
    \hat H = \int_{\mathbb{R}^3} d^3 x
    \left[ \hat \psi^\dagger \left( -\frac{\hbar^2}{2m}\nabla^2 + V \right) \hat \psi + \frac{g}{2}  \hat\psi^\dagger \hat\psi^\dagger \hat\psi \,  \hat\psi \right]
    \label{H_contact_interaction},
\end{eqnarray}
where $\hat\psi(\vec x)$ represents the annihilation operator for a  boson of mass $m$, $V(\vec x)$ describes an external trapping potential, and $g$ represents a coarse-grained contact interaction between particles. In terms of microscopic physics, $g=4\pi\hbar^2 a_s / m$ where $a_s$ is the s-wave scattering length \cite{Pethick_2008,Pitaevskii_2016,Andersen_2004,Yukalov_2004}. Here, we study the cases of no interactions $(g=0)$ and repulsive ones ($g>0$).

\begin{figure}[htp!]
  \includegraphics[width=\columnwidth]{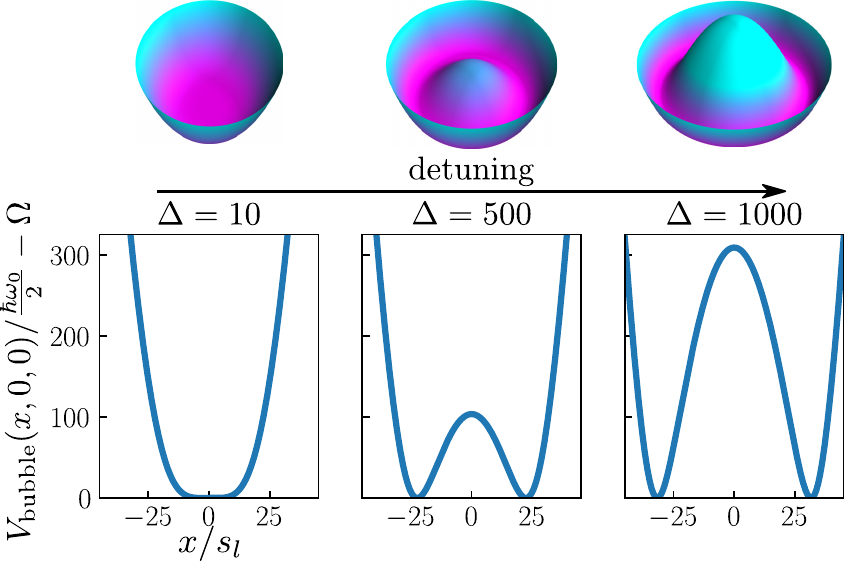}
  \caption{
    (Color online)
    Top (bottom) row: surface plots (spatial slices) of the bubble trap at various detuning values. As $\Delta$ increases, the location of the radially symmetric potential minimum, $s_l\sqrt{\Delta}$, expands outwards leading to the formation of shell-shaped structures.
  }
  \label{bubble_trap}
\end{figure}

First, we consider an idealized rotationally-symmetric ``bubble trap" potential which can tune between 3D spheres and quasi-2D thin spherical shells \cite{Zobay_2001,Zobay_2004}:
\begin{eqnarray}
    V_\text{bubble}(\vec x) = \frac{1}{2} m \omega_0^2 s_l^2 \sqrt{ \left[ \left( |\vec x| / s_l \right)^2 - \Delta  \right]^2 +(2\Omega)^2 }
    \label{V_bubble},
\end{eqnarray}
where $s_l=\sqrt{\hbar/2m\omega_0}$ is the oscillator length associated with frequency $\omega_0$ and $\Delta$ and $\Omega$ are dimensionless parameters that control the radius and width of the potential well \cite{Padavic_2017,Sun_2018}.
For instance, a positive $\Delta$ sets the potential minimum at a radius of $s_l \sqrt{\Delta}$.
By varying these parameters, we can probe thermodynamics through the range of desired geometries, as shown in Fig.~\ref{bubble_trap}.
Physically, $\Delta$ ($\Omega$) corresponds to the rf detuning (Rabi frequency) of a trapped atomic gas \cite{Lundblad_2019}.

When considering the bubble trap above, we work in units where $s_l=\omega_0=m=k_B=1$, and set all other parameters to correspond closely with the physically relevant case of ongoing CAL experiments. Namely, we take the number of atoms to be $N=50 000$, the dimensionless ``width" parameter to be $\Omega=250$, and (to expand from a filled-sphere to a hollow shell) consider a range of values for the dimensionless ``radius" parameter: $\Delta=0\to 1000$.
By fixing the particle number, this expansion process differs from \cite{Tononi_2019} where BECs on the surface of a sphere were investigated and expansions could be considered by varying the radius at fixed density.
We also note that negative detunings are relevant experimentally, but because the role of $\Delta$ in the bubble trap is to tune from filled-sphere to thin-shell geometries, we simply consider $\Delta \ge 0$.
For atoms in the internal state $|F=2,m_F=2\rangle$ subject to a trap with frequency $\omega_0/2\pi=80~\text{Hz}$, these values correspond to a Rabi frequency of $(\Omega \omega_0/ 4)/2\pi = 5~\text{kHz}$ and a maximum rf detuning of $(\Delta \omega_0/ 8)/2\pi = 10~\text{kHz}$.
In cases where we consider nonzero interactions, we use a value appropriate for both $^{87}\text{Rb}$ and the CAL experiment by setting the (dimensionless) interaction strength to $8\pi N a_s/ s_l = 7000$. For $N=5\times10^4$ atoms in an $80~\text{Hz}$ trap, this corresponds to a scattering length on the order of $5~\text{nm}$.

\subsection{Spectrum and states}

We first find the energy spectrum and eigenstates for a noninteracting system confined by the bubble trap of Eq.~(\ref{V_bubble}). Due to its rotational symmetry, eigenstate wavefunctions can be written in the form
\begin{eqnarray}
    \phi_{kl{m_l}}(\vec x)=\frac{1}{r}u_{k l}(r) Y_l^{m_l}(\theta,\phi)
    \label{single_particle_wavefns},
\end{eqnarray}
where $(r,\theta,\phi)$ are spherical coordinates, and $Y_l^{m_l}$ are spherical harmonics having $l=0,1,2,\dots$ and ${m_l}=-l,\dots,l$. The quantum number ``$k$" completes the indexing by accounting for the radial direction. Obtaining the eigenstates then reduces to solving the radial component of the Schr{\"o}dinger equation:
\begin{eqnarray}
    \left( -\frac{\hbar^2}{2m}\frac{d^2}{dr^2} + V(r) + \frac{\hbar^2 l(l+1)}{2m r^2} \right) u_{k l}(r) = \varepsilon_{kl} u_{k l}(r)
    \label{radial_eq},\quad
\end{eqnarray}
where $u_{k l}(0)=u_{k l}(\infty)=0$. For fixed angular momentum quantum number $l$, the index $k=0,1,2,\dots$ is chosen to correspond to increasing energy eigenvalues: $\varepsilon_{0,l} \le \varepsilon_{1,l} \le \varepsilon_{2,l} \le \cdots$.

\begin{figure}[htp!]
  \includegraphics[width=\columnwidth]{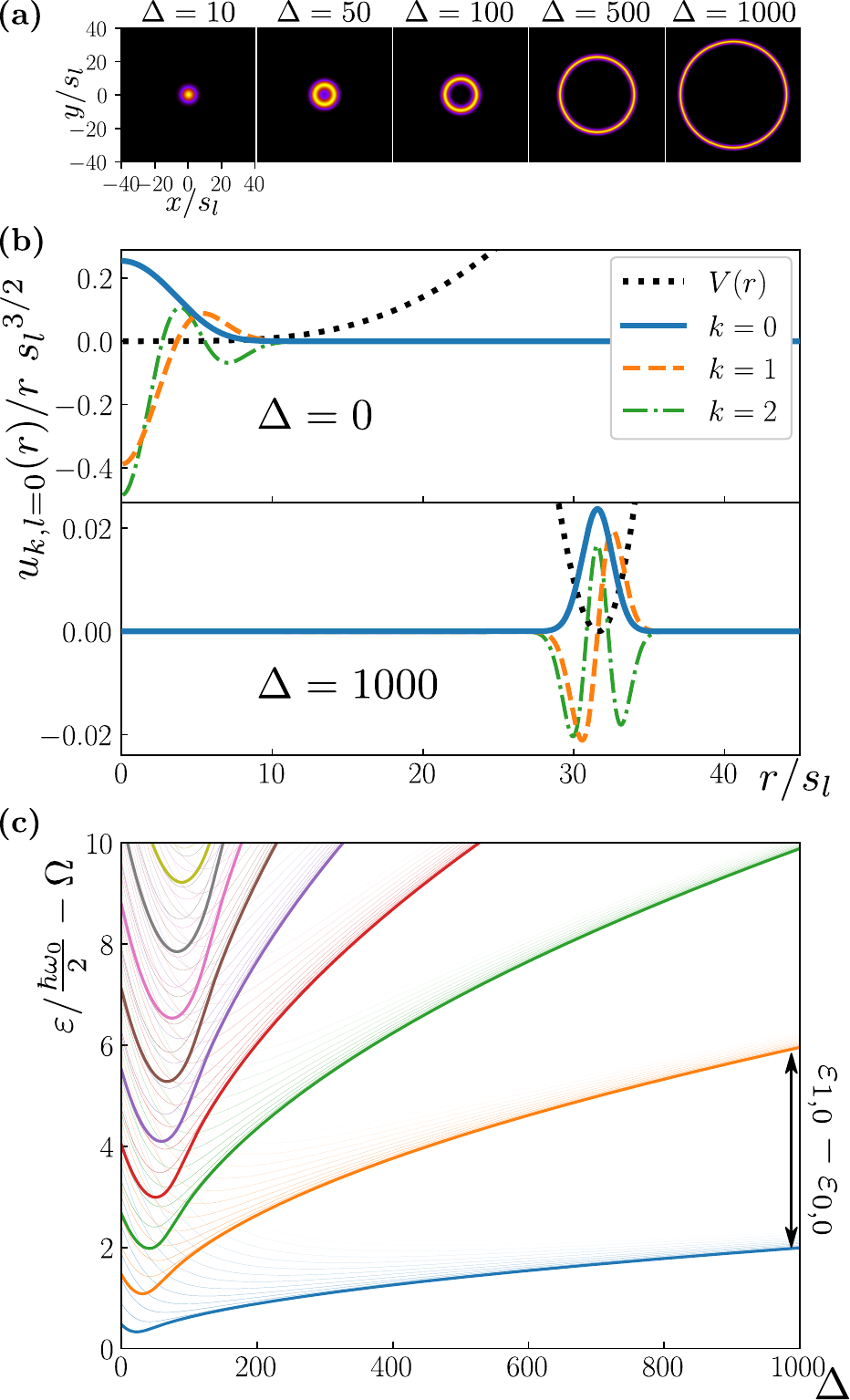}
  \caption{
    (Color online)
    Eigenstates and spectrum of the noninteracting bubble-trapped system
    (a)
    Probability density for the ground state wavefunction in the $z=0$ plane at various detuning values.
    In each plot, the color is normalized such that black (yellow) corresponds to the minimum (maximum) value.
    (b)
    Radial components of wavefunctions in the bubble trap, $u_{kl}(r)/r$. Only the zero angular momentum states for $k=0,1,2$ are shown (along with a shifted and scaled potential in the background). We show the wavefunctions in the two extreme limits of a filled-sphere, $\Delta=0$, and a thin-shell, $\Delta=1000$.
    (c)
    Partial spectrum of the bubble trap with the potential minimum, $\Omega \frac{\hbar\omega_0}{2}$, subtracted off.
    The color is based on the index $k$, with $k=0$ corresponding to blue, $k=1$ corresponding to orange, and so on. As $l$ increases, so does the corresponding energy level $\varepsilon_{kl}$ for a given $k$. Higher angular momentum states have been removed from the image for clarity.
  }
  \label{bubble_states}
\end{figure}

We employ a simple finite-difference method to discretize and then numerically solve Eq.~(\ref{radial_eq}) as the detuning parameter $\Delta$ is varied. As $\Delta$ increases (see Fig.~\ref{bubble_trap}), the peak of each eigenstate's probability density will begin to expand radially outward.
Hence, a state localized near the origin for small $\Delta$ becomes squeezed into a hollow thin shell at large $\Delta$. Fig.~\ref{bubble_states}a shows the evolution of the bubble trap ground state during this hollowing-out procedure.
As the bubble expands and the ground state eventually hollows-out, there is a change in topology characterized by different second homotopy groups: the filled-sphere, technically a $3$-ball, is contractible (i.e. a $2-$sphere inside this space can be continuously deformed to a point) whereas the hollow-sphere is not \cite{Padavic_dissertation_2020}.
However, because the ground state probability density decays gradually towards zero away from its peak, determining a precise $\Delta$ at which a change in topology occurs is subtle.
We note that another feature of the noninteracting ground state is that its peak density remains at the origin until some critical detuning value ($\Delta \approx 27$) upon which it begins to move radially outward.

In the case of a thin spherical shell, it is straightforward to approximate the geometry of the ground state.
In this limit, the bubble trap potential can be approximated by a shifted harmonic oscillator with frequency $\omega_0 \sqrt{\Delta/\Omega}$ and radial shift $s_l\sqrt{\Delta}$ corresponding to the bubble trap's radial potential minimum.
Using a one-dimensional shifted oscillator to capture the the radial behaviour of the ground state, we expect the thickness of the shell to scale as $s_l (\Omega / \Delta)^{1/4}$. This means the approximate volume of the ground state density scales like $s_l^3 \Delta^{3/4} \Omega^{1/4}$. We see, therefore, that as the bubble expands with increasing $\Delta$ and fixed $\Omega$, the thickness of the shell decreases but the total volume increases.

Turning our attention to the behavior of the radial component of the wavefunctions $u_{kl}(r) / r$ (Fig.~\ref{bubble_states}b), we note that the quantum number $k$ labels the number of nodes in each radial wavefunction (disregarding zeros due to boundary conditions). Thus, for fixed angular momentum $l$, increasing $k$ corresponds both to larger energy eigenvalues and node count.
At large $\Delta$, $k=0$ states, which have no radial nodes, compose a low-lying energy band (Fig.~\ref{bubble_states}c) and constitute a quasi-2D basis.

A partial spectrum of the bubble trap is shown in Fig.~\ref{bubble_states}c.
Note that for a given $k$ value, as the angular momentum quantum number $l$ increases, so does the corresponding energy $\varepsilon_{kl}$.
Furthermore, for a given $l$ value, we find the energy difference between adjacent $k$ bands begins to increase with the detuning parameter $\Delta$.
The spacing between the bottom of the $k=0$ and $k=1$ bands represents the energy scale at which the system enters a quasi-2D regime.

\section{Bubble trap thermodynamics (noninteracting)}

In this section, we use the above bubble-trapped states to numerically compute thermodynamic quantities in the absence of interactions as the trap evolves from a filled sphere to thin shell geometry. We find that adiabatic expansion of a BEC leads to condensate depletion and hence an initial condensate can be lost during the process.

\subsection{BEC critical temperature}

\begin{figure}[htp!]
  \includegraphics[width=\columnwidth]{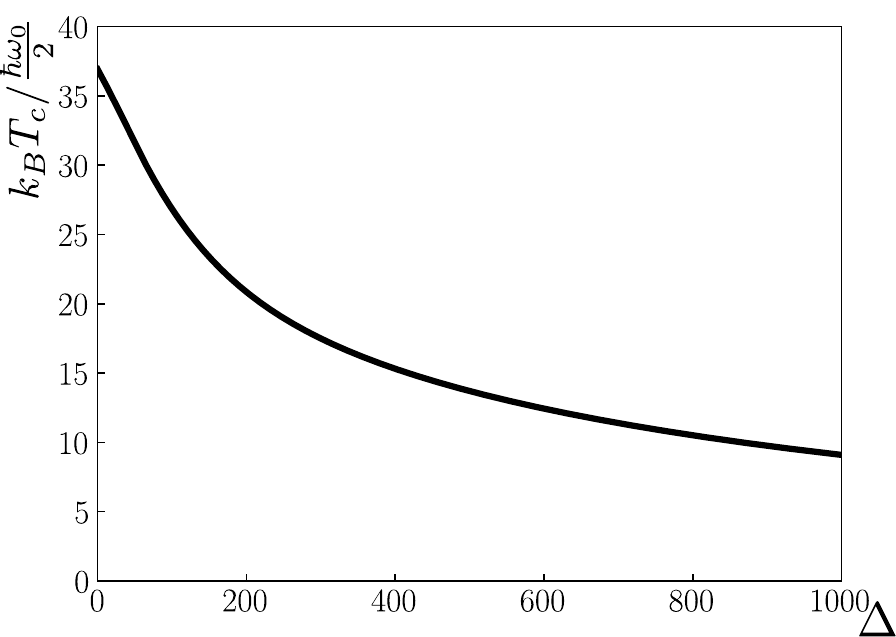}
  \caption{
    The BEC critical temperature of $N=5\times10^4$ bosons in the bubble trap versus the detuning parameter $\Delta$, which controls the shell thickness.
    $\Delta=0\to 1000$ corresponds to continuously deforming the system from a filled-sphere to hollow thin-shell (Fig.~\ref{bubble_trap}).
    The temperature is given in units of $\frac{\hbar\omega_0}{2 k_B}$ where $\omega_0$ is the oscillator frequency in Eq.~(\ref{V_bubble}).
  }
  \label{bubble_Tc}
\end{figure}

For a noninteracting system, thermodynamic properties follow from Eq.~(\ref{radial_eq}) (see appendix \ref{thermo_appendix} for details on the thermodynamic formalism). Of crucial importance for shell-shaped condensates is the BEC critical temperature, $T_c$, which is computed by solving the implicit equation:
\begin{eqnarray}
    N = \sum_{kl \ne 0} (2l+1) \frac{1}{e^{(\varepsilon_{kl}-\varepsilon_0)/ k_B T_c }-1}
    \label{Tc_implicit}.
\end{eqnarray}
where $N$ is the number of particles in the system and we denote the single-particle ground state ($k=0$, $l=0$, ${m_l}=0$) with subscript $``0"$.
From the spectrum of the bubble trap, we can determine the BEC critical temperature as a function of detuning, $\Delta$. Fig.~\ref{bubble_Tc} shows that as the system expands from a filled sphere to a hollow, thin shell, the critical temperature drops significantly.

Qualitatively, this can be explained by considering the phase-space density $n \lambda_T^{3}$ where $n=N/\text{Vol}$ is the particle density with $\text{Vol}$ representing the characteristic volume of the bubble and $\lambda_T \equiv (2\pi \hbar^2 / m k_B T)^{1/2}$ is the thermal de Broglie wavelength.
Condensation should occur when the phase-space density is on the order of unity.
As the system expands from a filled sphere into a thin shell, the volume of the shell increases while the particle number is fixed, hence the particle density decreases. Therefore, one must lower the temperature to obtain a phase-space density on the order of unity.

In addition to the reduction of $T_c$ at large detuning, the function $T_c(\Delta)$ decreases monotonically with $\Delta$ and also possesses an inflection point at a small detuning value of $\Delta \approx 38$.
With regard to our earlier discussion on the noninteracting ground state of the bubble trap, this inflection point occurs slightly after the peak density of the wavefunction begins to move radially outward, but before the ground state becomes hollow. For $\Delta = 38$, the density of the ground state wavefunction at the origin is approximately $62\%$ of its peak value and hence clearly has yet to hollow-out.

In contrast to a recent calculation by Tononi et al \cite{Tononi_2020} of the BEC critical temperature of a Bose gas in a similar geometry, which relied on the semiclassical approximation, here we obtain our result using the (numeric) spectrum of the bubble trap. Shortly, we will directly compare the semiclassical method with the spectral method and find that the $T_c$ obtained from the spectrum is lower than that obtained semiclassically.

\subsection{Adiabatic expansion and cooling}

To model the creation of large bubbles on CAL, we consider an initially harmonic trap (and filled, spherical system) being expanded outwards by increasing the detuning parameter $\Delta$. We assume this process is performed adiabatically (i.e. without generating heat) and at fixed particle number.
This approach is consistent with experimental CAL results reported in \cite{Lundblad_2021} where the trapping potential was dynamically deformed sufficiently slow to maintain adiabaticity.
Here, we enforce adiabaticity by considering an isentropic (constant entropy) process and compute the temperature during adiabatic expansion by solving the implicit equations for net particle number $N$ and entropy $S$:
\begin{subequations}
    \label{NI_thermodynamics}
    \begin{eqnarray}
        N &&= \sum_{kl} (2l+1) f_{kl}
        \label{NI_N},
        \\
        S &&= k_B \sum_{kl} (2l+1) \left[ (1+f_{kl})\ln(1+f_{kl}) - f_{kl} \ln f_{kl} \right]
        \label{NI_S},\quad\quad
    \end{eqnarray}
\end{subequations}
where $f_{kl}=1/(e^{\beta(\varepsilon_{kl}-\mu)}-1)$ is the Bose-Einstein distribution function at chemical potential $\mu$ and inverse temperature $\beta(=1/k_B T)$. We model the adiabatic expansion process as follows: we fix a starting temperature (prior to expansion, at $\Delta=0$) and then solve Eq.~(\ref{NI_thermodynamics}) to find the associated entropy. Next, we model expanding the gas by increasing $\Delta$. At each stage of the expansion, we determine the new temperature by solving Eq.~(\ref{NI_thermodynamics}) such that both the particle number $N$ and the entropy $S$ are fixed.
Note that when $T>T_c$ one must solve for both temperature and chemical potential, but when $T<T_c$ and $\mu\to\varepsilon_0^-$, one must instead solve for the temperature and number of condensed particles, $N_0$.

We note that one expects adiabatic expansion of the bubble cools the system. For instance, during an isentropic process, a free or harmonically trapped BEC obeys the characteristic relation $\text{Vol} \, T^{\nu} = \text{const.}$ where the exponent $\nu$ is positive; hence, increasing the volume of the gas results in a decrease in temperature.

\subsection{Cooling by adiabatic expansion depletes the condensate}

\begin{figure}[htp!]
  \includegraphics[width=\columnwidth]{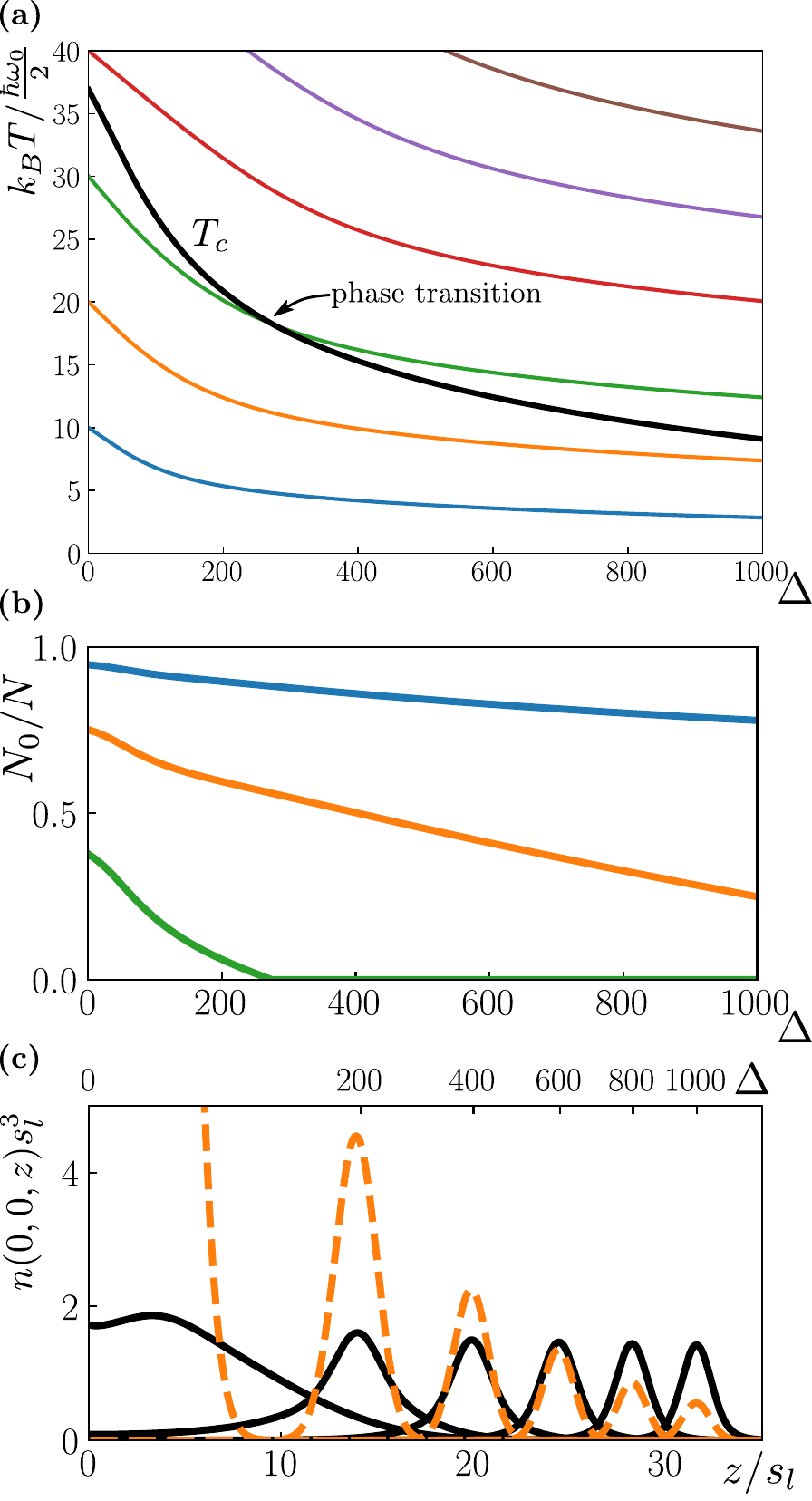}
  \caption{
    (Color online)
    (a)
    BEC critical temperature (black line) of $N=5\times10^4$ particles in the bubble trap along with various adiabatic expansion temperature profiles (colored lines). Each adiabatic expansion corresponds to a different initial temperature when $\Delta=0$: $k_B T_i / \frac{\hbar \omega_0}{2} = 10, 20, 30, \dots$.
    Note that $T_c$ decreases faster than any adiabatic temperature profile; thus, expanding the gas isentropically leads to a decreasing condensate fraction.
    (b)
    Plot of the condensate fraction, $N_0/N$, during the adiabatic expansions shown in the previous image.
    (c)
    Comparison of the local condensate density (colored-dashed lines) and excited state density (black-solid lines) at various stages of the $k_B T_i = 20 \frac{\hbar\omega_0}{2}$ adiabatic expansion (shown along the $z$-axis).
    Above each density curve is the associated value of $\Delta$ (marked at the potential minimum location $s_l \sqrt{\Delta}$).
  }
  \label{bubble_adiabatic_expansion}
\end{figure}

We next present numerical solutions for the temperature of a bubble-trapped system during adiabatic expansion at various starting temperatures; the results are summarized in Fig.~\ref{bubble_adiabatic_expansion}a.
We note that the BEC critical temperature decreases faster than the temperature of the gas as it adiabatically expands.
This means that if one does not sufficiently cool the system before beginning to expand it into a bubble shape, an initially condensed system transitions during expansion into a thermal gas.
An explicit example of this can be seen for the adiabatic expansion temperature profile with initial temperature $k_B T_i = 30 \frac{\hbar \omega_0}{2}$ (which corresponds roughly to a temperature of $60~\text{nK}$ for a $80~\text{Hz}$ trap). Although the system before expansion is in the condensed phase, it eventually finds itself above the condensate transition temperature as the parameter $\Delta$ increases.

Even if the initial temperature is low enough to remain in the condensed phase for the entire expansion, we find that the condensate fraction decreases as the system expands.
In other words, adiabatic expansion in this case leads to a loss of phase-space density; this is a negative version of the long-known phase-space density increases at constant entropy exploited in various experiments \cite{Pinkse_1997,Stamper-Kurn_1998,Weber_2003,Lin_2009}.
Figure.~\ref{bubble_adiabatic_expansion}b explicitly shows the depletion upon adiabatic expansion for a few different initial temperatures by plotting the condensate fraction $N_0/N$.
This effect is pronounced when comparing the local condensate density with the excited state density at various stages in the expansion process.
If we consider a condensed gas being expanded adiabatically whose initial temperature is $k_B T_i = 20 \frac{\hbar \omega_0}{2}$, we show in Fig.~\ref{bubble_adiabatic_expansion}c that there is clear condensate depletion at large $\Delta$.

It is important to note that although condensate depletion during adiabatic expansion presents an experimental hindrance to observing large BEC shells, Fig.~\ref{bubble_adiabatic_expansion}b also shows that the degree of severity depends crucially on the initial temperature prior to expansion. A system which starts off colder experiences depletion at a far slower rate when compared to one which starts off warmer.

\subsection{Validity of the semiclassical approximation}

\begin{figure}[htp!]
  \includegraphics[width=\columnwidth]{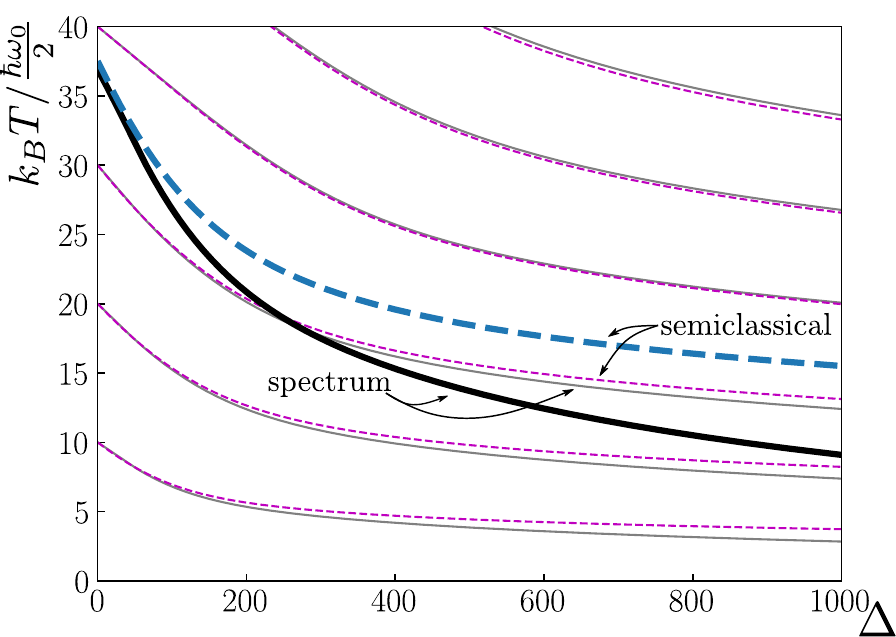}
  \caption{
    (Color online)
    Comparison of our previous results using the spectrum of the bubble trap with the semiclassical approximation.
    Here, we show the BEC critical temperature for $N=5\times10^4$ particles along with various adiabatic expansion temperature profiles.
    [The undashed (dashed) black (blue) line is $T_c$ using the spectral (semiclassical) approach. The undashed (dashed) grey (magenta) lines are adiabatic expansion profiles using the spectral (semiclassical) approach.]
  }
  \label{semiclassical_comparison}
\end{figure}

The semiclassical approximation (see appendix \ref{thermo_appendix}) is a standard technique which allows one to perform calculations with arbitrary trapping potentials. It is thus instructive to see how results obtained using this method compare with those obtained using the (numeric) bubble trap spectrum (Fig.~\ref{semiclassical_comparison}).
Importantly, we find that the predictions of the semiclassical approximation become less accurate as the shell is expanded (in particular, note the difference in predicted BEC critical temperature at large $\Delta$).
In order to explain these discrepancies, we note that the semiclassical approximation assumes the system is large enough to ignore boundary effects and treat momentum as continuous. During expansions however, this assumption is eventually violated in the radial direction as the shell thickness becomes vanishingly small.
From Fig.~\ref{semiclassical_comparison}, we conclude that semiclassical predictions should be taken with some level of caution for thin-shell systems or fully-expanded bubbles.

Notably, the semiclassical prediction for $T_c$ in the thin-shell becomes considerably higher than that obtained using the spectrum.
At the highest detuning shown, the difference is about $6.5 \frac{\hbar\omega_0}{2}$ (roughly $12.5~\text{nK}$ for a $80~\text{Hz}$ trap). Hence, for some range of initial temperatures before expansion, the semiclassical approximation will incorrectly predict that an expanding gas will remain condensed when in fact it will transition to the normal phase.

Finally, we note that when $T \gtrsim T_c$, the temperature predictions during adiabatic expansion don't vary greatly between the spectral and semiclassical methods. Furthermore, even below $T_c$, the difference between the two methods is far less severe than that seen in the prediction of the BEC critical temperature.

\section{Effects of interactions and dimensionality}

Below, the inclusion of interactions is discussed as the bubble trap potential is modified to expand a filled sphere condensate into a thin shell.
At $T=0$, if the particle number is held fixed, we find a noninteracting description to be a good approximation for thin shells.
We then use this result to develop an effective low temperature theory in a spontaneous $U(1)$ symmetry broken phase and find, in the thin-shell regime, that interactions do little to modify the results obtained previously.
We conclude by considering the breakdown of this mean-field theory approach as one crosses over from 3D to 2D physics.

\subsection{Validity of the noninteracting description}

Having described the thermodynamics of a bubble trapped gas in the noninteracting ($g=0$) limit, we now address the question of where the noninteracting description accurately captures salient features even when interactions are present.
Working at zero temperature, we consider the evolution of a condensate in the bubble trap as the detuning parameter increases and show that for thin shells at fixed particle number, the noninteracting picture indeed constitutes a reasonable description.
We justify this assertion in two ways: 1) we analyze numeric solutions of the Gross-Pitaevskii equation for an experimentally-relevant interaction strength and contrast this data with both
weakly-interacting and interaction-dominated regimes
of the Gross-Pitaevskii equation, 2) we perform a variational calculation to gain insight into the role of interactions in the thin-shell limit.

{\bf Comparing condensate wavefunctions}:
The many-body ground state of Eq.~(\ref{H_contact_interaction}) is characterized by a condensate wavefunction $\psi_c(\vec x)$ which obeys the time-independent Gross-Pitaevskii (GP) equation:
\begin{eqnarray}
    0
    = \left( - \frac{\hbar^2}{2m}\nabla^2 + V(\vec x) - \mu + g |\psi_c(\vec x)|^2 \right) \psi_c(\vec x)
    \label{GP}.
\end{eqnarray}
and is normalized as $\int d^3x |\psi_c|^2 \sim N$  as $T\to 0$.
Because the GP equation is nonlinear, analytic solutions are typically unfeasible.
Fortunately, in cases which correspond physically to non-interacting and interaction-dominated regimes of the GP equation, analytic solutions can be found.
When the effect of interactions can be considered small relative to the kinetic and potential terms in Eq.~(\ref{GP}), one can take the $g\to0$ limit of the GP equation and denote the corresponding solution as the noninteracting (NI) condensate wavefunction: $\psi_\text{NI}(\vec x) = \sqrt{N_0} \phi_0 (\vec x)$ where the subscript ``$0$" denotes the ground state of Eq.~(\ref{radial_eq}).
In the opposite limit, when the interaction term dominates the GP equation, we use the Thomas-Fermi (TF) approximation which treats the kinetic contribution to Eq.~(\ref{GP}) as negligibly small \cite{Pethick_2008,Pitaevskii_2016}.
Dropping the kinetic term allows one to solve the GP equation to obtain the TF condensate wavefunction: $|\psi_\text{TF}(\vec x)|=\sqrt{(\mu-V(\vec x))/g}$ when $V(\vec x) < \mu$ and $\psi_\text{TF}=0$ otherwise \cite{Pethick_2008,Pitaevskii_2016}.

Numerical solutions for the ground state of Eq.~(\ref{GP}) (at a range of $\Delta$ values for the bubble trap potential, Eq.~(\ref{V_bubble})) were found using an imaginary-time algorithm~\cite{Chiofalo_2000} and taking the experimentally-relevant value of dimensionless interaction strength, $8\pi N a_s/s_l=7000$.

\begin{figure}[htp!]
  \includegraphics[width=\columnwidth]{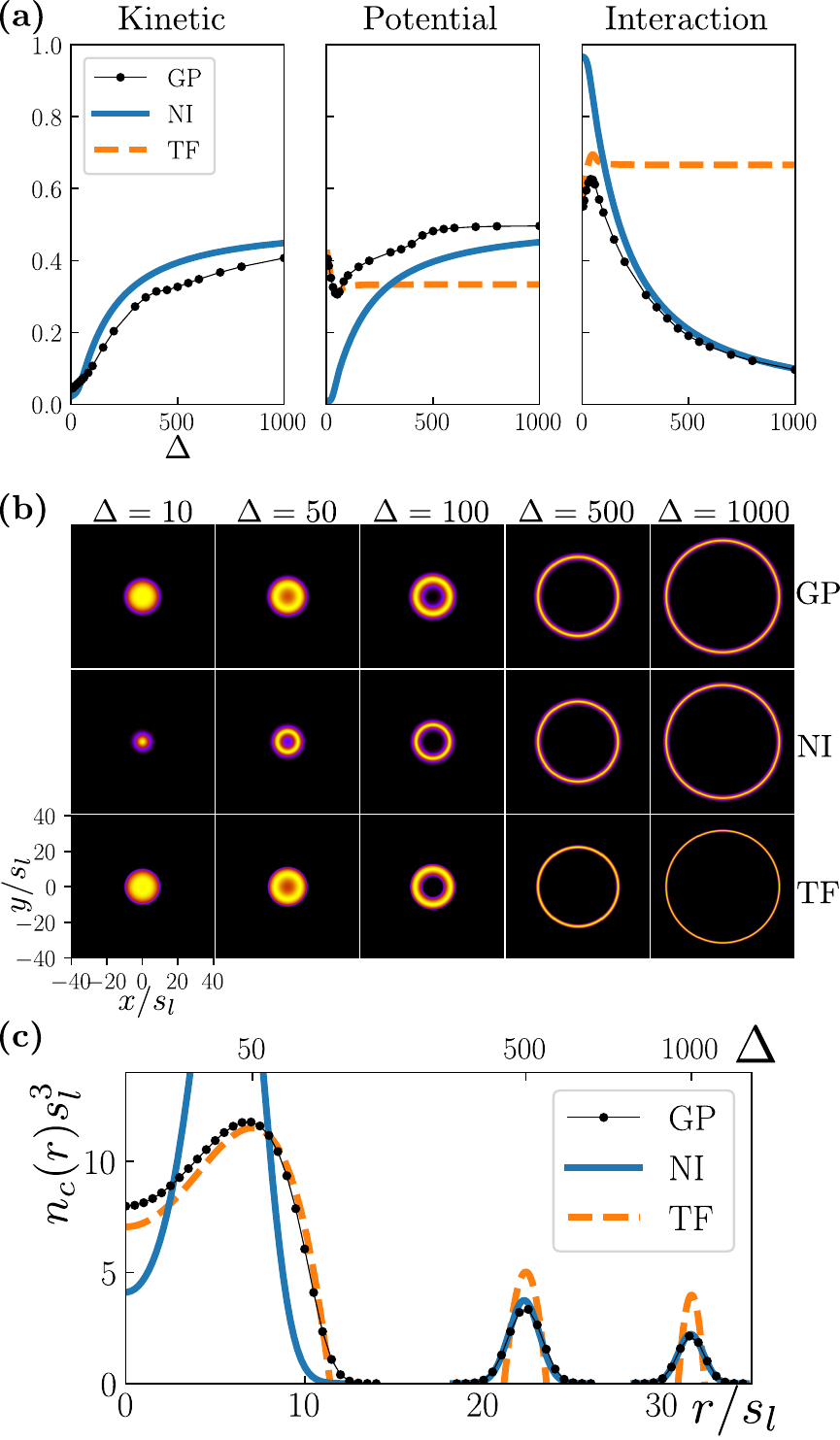}
  \caption{
    (Color online)
    Comparison between the zero-temperature Gross-Pitaevskii (GP), noninteracting (NI), and Thomas-Fermi (TF) condensate wavefunctions at various detunings for $N=5\times10^4$ atoms in the bubble trap.
    (a)
    From left to right, the fraction of energy stored as kinetic, trap potential, and interaction energy. Note: the bubble trap reference energy has been subtracted off from the potential. We also treated quantum depletion as nearly negligible for the NI condensate wavefunction \cite{Pethick_2008} (we set $N_0 = 0.99 N$ for $T = 0$).
    (b)
    Local condensate density at various detunings. The density is plotted in the $z=0$ plane: $n_c(x,y,0)=|\psi_c(x,y,0)|^2$. In each plot, the color is normalized such that black (yellow) corresponds to the minimum (maximum) density value.
    (c)
    Local condensate density plotted against the radial coordinate for the detunings $\Delta=50,500,1000$ displayed in (b).
  }
  \label{condensate_wavefn_comparison}
\end{figure}

We now calculate and compare the kinetic energy, $\int d^3 x \psi^*( - \frac{\hbar^2}{2m} \nabla^2 ) \psi$, trap potential energy, $\int d^3 x V |\psi|^2$, and interaction energy, $\int d^3 x \frac{g}{2} |\psi|^4$, at zero-temperature using the numerically-solved GP ($\psi\to\psi_c$), NI ($\psi\to\psi_\text{NI}$), and TF ($\psi\to\psi_\text{TF}$) forms of the condensate wavefunctions.
Consistent with the TF approximation, we treat the kinetic energy of $\psi_\text{TF}$ as negligibly small in comparison to the energy stored in interactions.
In the case of the NI condensate wavefunction, its smooth spatial profile allows one to insert it into each energy functional (including the interaction energy).
However, in order for the NI approximation to be self-consistent, the interaction energy associated with $\psi_\text{NI}$ must be small compared to kinetic and potential energy contributions.
Fig.~\ref{condensate_wavefn_comparison}a shows the energy fractions as $\Delta$ is increased (expanding the system from a filled-sphere to a thin-shell) at $T=0$, with fixed particle number, using the three wavefunctions.

From the numerically-solved GP results, one can see clearly that the energy stored in interactions is the dominant contribution for small $\Delta$ (filled spheres), but becomes the least significant contribution for large $\Delta$ (thin shells).
For parameter regimes in which the interaction energy fraction is sufficiently suppressed, one expects the NI case to give a reasonable description of the system.

Let us now compare results between the various condensate wavefunctions.
For small $\Delta$, the TF approximation does reasonably well compared to the numerical solution (GP) whereas the NI approximation does poorly; hence, interactions are expected to be important in the description of a bubble-trapped system in this parameter regime.
Conversely, as $\Delta$ increases, and the shell expands, the TF approximation becomes a poor approximation of the GP numerics, whereas the NI approximation performs reasonably well.
In particular, the TF energy fractions become nearly constant, whereas the GP interaction energy fraction falls drastically and begins to converge with the NI wavefunction results for $\Delta\gtrsim 300$.
Thus, for thin shells, we expect a NI description is a good approximation to the full solution with interactions.
The breakdown of the TF approximation as $\Delta$ increases is due to its failure to capture the rising kinetic energy fraction of the GP condensate in this regime.
We also note that, although the NI and GP interaction energy fractions begin to converge, the NI wavefunction systematically overestimates the kinetic and underestimates the trap potential energy fractions.
To further contextualize these conclusions, we direct the reader to Fig.~\ref{condensate_wavefn_comparison}b and \ref{condensate_wavefn_comparison}c which show the ($T=0$) local condensate density at various detuning values.
Here, one can see clearly that for filled spheres (thin shells) the TF (NI) condensate wavefunction captures the salient spatial features of the GP results.

{\bf Variational calculation}:
To further understand the thin shell results, let us look for an origin to the decreasing interaction energy fraction during the zero-temperature expansion.
One might suspect interactions play a prominent role when the shell becomes thin and atoms get squeezed into a space of small width.
However, one should keep in mind that the particle density reduces as the system becomes thinner: the particle number is held fixed, while the condensate volume, which we argued earlier should scale as $s_l^3 \Delta^{3/4} \Omega^{1/4}$ in the NI case, is increasing as the bubble expands.

To see this more concretely, we perform a variational calculation \cite{Pethick_2008} with a trial condensate wavefunction:
\begin{eqnarray}
    \psi_\text{trial}(\vec x) = A  \, F\left( \frac{|\vec x|-a}{b} \right) e^{i\varphi(\vec x)}
    \label{psi_trial},
\end{eqnarray}
where the variational parameters $A$, $a$, and $b$ are all nonnegative real constants, $\varphi(\vec x)$ is the phase of the condensate wavefunction, and $F(\cdot)$ is a dimensionless (nonnegative) smooth real function which decays to zero for large arguments \cite{Lannert_2007}. For example, we could use a Gaussian, $F(u)=e^{-u^2 / 2}$ \cite{Pethick_2008,Lannert_2007}, from which one can see the length scales have a natural interpretation: $a\approx (R_\text{out} + R_\text{in})/2$ describes the average shell radius and $b \approx R_\text{out} - R_\text{in}$ describes the shell thickness where $R_\text{out}$ ($R_\text{in}$) is the outer (inner) radius of the condensate shell. In this context, we can define a thin-shell as one for which $a/b \gg 1$.

We now insert the trial condensate wavefunction into $H[\psi^*,\psi] - \mu N[\psi^*,\psi]$ where $H[\cdot]$ and $N[\cdot]$ are the Hamiltonian and particle number functionals respectively.
Considering no superflow, $\frac{\hbar}{m} \nabla \varphi = 0$, we extremize the functional and obtain the following forms for the kinetic, trap potential, and interaction energy terms in the thin shell limit, $b/a\to 0$:
\begin{subequations}
    \label{E_varitational}
    \begin{eqnarray}
        E^\text{trial}_\text{kin} && \equiv
        \int_{\mathbb{R}^3} d^3 x \, \psi_\text{trial}^*\left( - \frac{\hbar^2}{2m} \nabla^2 \right) \psi_\text{trial}
        \sim C_1 \frac{N}{b^2}
        \label{Ekin_variational},
        \\
        E^\text{trial}_\text{pot} && \equiv
        \int_{\mathbb{R}^3} d^3 x \, V |\psi_\text{trial}|^2
        \sim C_2 \, a^2 b^2 N
        \sim C_2 ' \, \omega_\text{sh}^2 b^2 N
        \label{Epot_variational},\qquad
        \\
        E^\text{trial}_\text{int} && \equiv
        \int_{\mathbb{R}^3} d^3 x \, \frac{g}{2} |\psi_\text{trial}|^4
        \sim C_3 \frac{N^2}{a^2 b}
        \label{Eint_variational},
    \end{eqnarray}
\end{subequations}
where we used the relationship $N \sim C_0 A^2 a^2 b$ as $b/a\to 0$ to replace the (amplitude) variational parameter $A$ with the particle number $N$ (which normalizes the trial wavefunction),
we introduced coefficients $C_0, \dots, C_3$ (and $C_2'$) which do not change as the system is expanded,
$\omega_\text{sh}\equiv\omega_0 \sqrt{\Delta/\Omega}$ is the frequency of the effective shifted harmonic oscillator one obtains in the thin shell limit of the bubble trap,
and in order to reach Eq.~(\ref{Epot_variational}), we took the thin shell limit such that $\Omega$ remained fixed and assumed the variational parameter $a$ was close to the location of the bubble trap's potential minimum: $a \sim s_l\sqrt{\Delta}$.
For clarity, in Eq.~(\ref{Epot_variational}) we also subtracted off the contribution due to the bubble trap reference energy $\Omega \frac{\hbar\omega_0}{2}$.

From these asymptotic results, one finds that the fraction of energy stored in interactions relative to the total energy becomes vanishingly small in the thin-shell limit.
This occurs because the ratio of interaction energy over kinetic energy is proportional to $N b /a^2$.
Therefore, for fixed particle number, the zero-temperature interaction energy fraction tends toward zero for increasingly thin shells.
Using this result we can further simplify our expression for the trap potential energy by recalling that in the noninteracting system the thickness of a thin shell scales as $s_l (\Omega / \Delta)^{1/4}$. By substituting this into Eq.~(\ref{Epot_variational}), one finds the trap potential energy, just as the kinetic energy, scales as $N / b^2$.

\subsection{Do interactions help preserve the condensate?}

In the NI case, we found earlier that adiabatically expanding a condensate into a thin shell led to a decreasing condensate fraction.
A natural question is how interactions modify this picture. However, based on the arguments of the previous section, we should not expect dramatic changes in the thin-shell limit where the NI case is expected to be a good approximation.
In general, computing thermodynamics in the presence of interactions is far more involved than in their absence \cite{Andersen_2004,Yukalov_2004}; here we proceed using a standard Bogoliubov quasiparticle description (see appendix \ref{thermo_appendix}).
In particular, at temperatures far below the $U(1)$ symmetry breaking transition, we develop a mean-field theory by expanding the boson field about the condensate wavefunction $\psi_c(\vec x)$. Importantly, we assume that the NI condensate wavefunction is a reasonable approximation to the true solution of the GP equation, $\psi_c(\vec x) \approx \psi_\text{NI}(\vec x)$, which we have shown above to be correct for a thin shell.
Thermodynamics then follow from solutions to the Bogoliubov equations (see appendix \ref{BOGO_bubble_appendix} for the case of the bubble trap) which we solve numerically using a finite-difference method.

As a means to probe the effect of interactions on the thermodynamics of an expanding system, we set a condensate fraction, $N_0/N$, and then model expanding the gas by increasing $\Delta$. Instead of doing the expansion adiabatically, here we simply ask: what temperature is required to maintain the given condensate fraction? This is motivated by the fact that the NI condensate wavefunction is a poor approximation of the true condensate wavefunction at low-detunings.
To evolve an initial ($\Delta=0$) temperature isentropically, we need to determine and fix the entropy at the start of the expansion, but because our effective theory assumes the NI condensate wavefunction is a reasonable reference state, any thermodynamic predictions we make at such small $\Delta$ are suspect.

\begin{figure}[htp!]
  \includegraphics[width=\columnwidth]{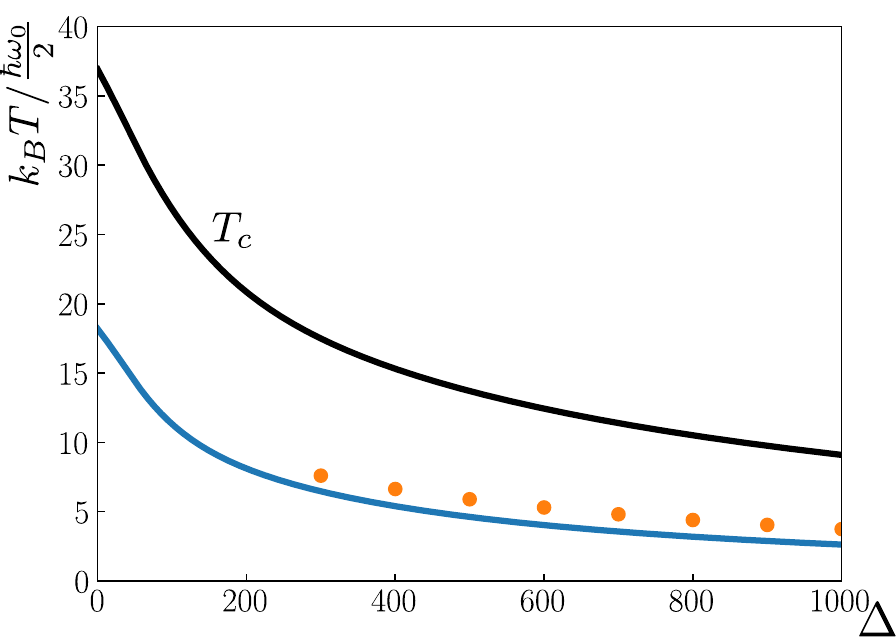}
  \caption{
    (Color online)
    Bubble trap expansions of $N=5\times10^4$ atoms done at a fixed-condensate fraction $N_0/N$. The black line shows the noninteracting (NI) BEC critical temperature, whereas the colored lines show the temperature required for an $80\%$ condensate fraction. The solid lines correspond to NI data whereas the markers correspond to Bogoliubov collective excitations.
    Because this Bogoliubov formalism assumes the NI condensate wavefunction is a reasonable approximation to the GP equation, we only show data points for $\Delta\ge 300$ where this assumption can be justified.
  }
  \label{BOGO_T_condensate_fraction}
\end{figure}

In Fig.~\ref{BOGO_T_condensate_fraction} we compare the temperatures required for an $80\%$ condensate fraction in the NI and Bogoliubov formalisms.
In the regime of applicability of our Bogoliubov description using the NI condensate wavefunction,
$\Delta\gtrsim 300$, we see that over this range of geometries interactions do increase the temperature required to obtain the given condensate fraction, but that this effect is very small. Consistent with the arguments of the the previous section, we thus find that interactions do little to change the thermodynamics of thin shells.
In a future work, we intend to investigate this question over the entire range of geometries afforded by the bubble trap by using the numerically-solved GP condensate wavefunction as the mean-field.

\subsection{Dimensional crossover}

While the previous thermodynamic calculations depend on mean-field considerations,
there are several limits in which these approaches would not suffice. First, the finite size of the sphere results in deviations from the thermodynamic limit. In principle, first order corrections could be taken into account using finite-size scaling analyses \cite{Goldenfeld_1992}.

As we consider thinner shells, a more drastic effect comes from reduction in effective dimensionality. The system enters the quasi-two dimensional regime at temperatures below which excitations along the radial direction become energetically unfavorable. More precisely, appealing to the energy spectrum in Fig.~\ref{bubble_states}c, energy levels $\varepsilon_{kl}$ are characterized by excitations along the radial ($k$) and angular ($l$) directions. For temperatures low enough to excite only the $k=0$ band, we have the condition $k_B T < \varepsilon_{1,0} - \varepsilon_{0,0}$. At the highest detuning shown in Fig.~\ref{bubble_states}c, we find $\varepsilon_{1,0} - \varepsilon_{0,0} \approx 4 \frac{\hbar\omega_0}{2}$ which corresponds to a temperature scale on the order of $10~\text{nK}$ for a $80~\text{Hz}$ trap. Below this temperature, and at the largest detunings, we expect the emergence of two-dimensional physics; this regime clearly requires treatments that go beyond mean-field.
As has been demonstrated in planar settings, the cross-over from effective three dimensional to quasi-two dimensional physics is particularly rich in the transition into Bose-Einstein condensed states \cite{Chomaz_2015}.

As with disk-shaped geometries, defects such as vortices form the natural source of excitations that complement the mean-field hydrodynamic long-wavelength, low-frequency excitations \cite{Fetter_2009}.
In previous work by two of the current authors and collaborators, we explored the physics of a vortex-antivortex pair in a thin spherical condensate shell and found the energy of the pair scales with shell thickness \cite{Padavic_2020}.
Furthermore, explorations in disk-shaped quasi-2D trapped gases show there is a temperature regime in which vortex-antivortex pairs appear \cite{Simula_2005,Choi_2013}.
Thus, as the bubble-trapped system hollows out, depending on the parameters, even before entering the two-dimensional limit, we may well access a finite-thickness regime in which vortex-antivortex pair excitations become favorable, destroying condensation in the system.

In the uniform two dimensional limit, the proliferation of such vortex-antivortex pairs renders the critical temperature to take the Berezinskii-Kosterlitz-Thouless (BKT) form \cite{Berezinskii_1972,Kosterlitz_1973}. The effect of curvature on such physics is highly interesting in and of itself \cite{Vitelli_2006,Turner_2010,Prestipino_2019,Moller_2020,Prestipino_2021,Bereta_2021}. In recent work, Tononi and co-workers demonstrated that in the precise shell setting considered here and of relevance to CAL, the BKT transition finds its finite-size manifestation \cite{Tononi_2021}.

\section{Applications to the CAL trap}

\begin{figure*}[htp!]
  \includegraphics[width=\textwidth]{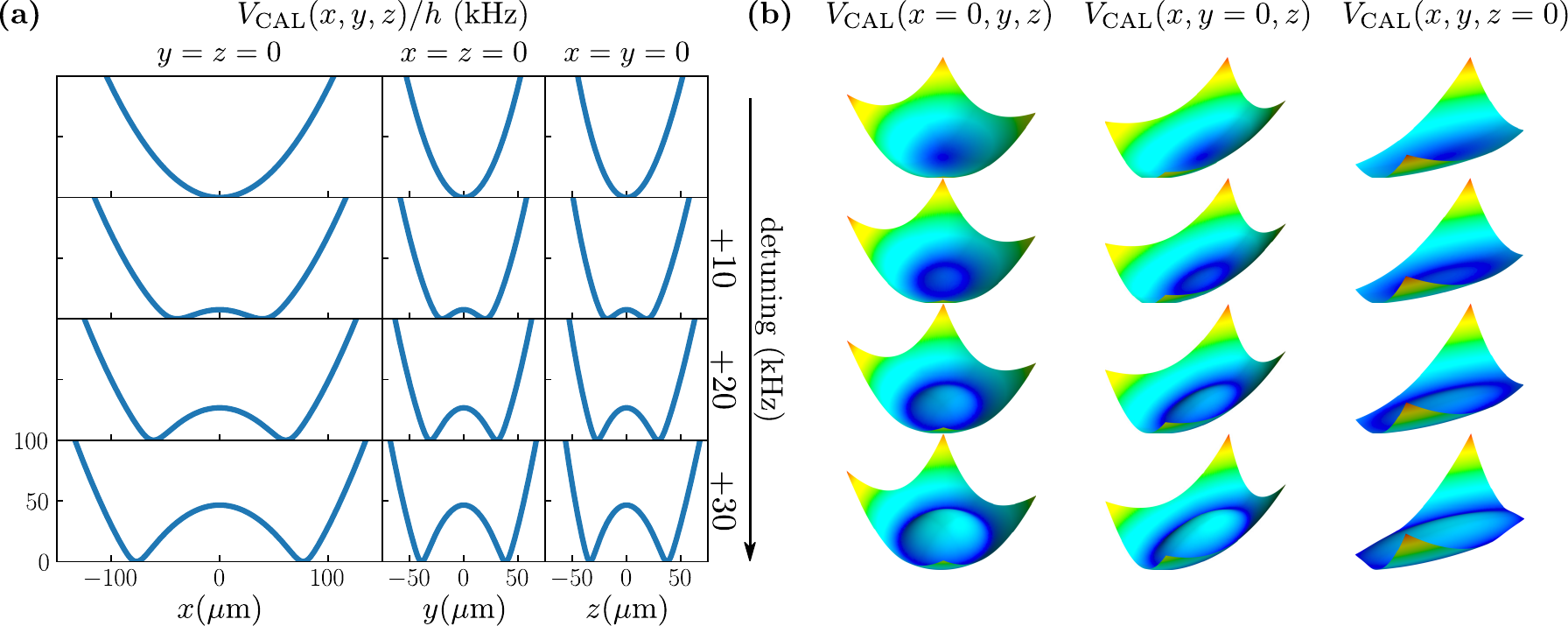}
  \caption{
    (Color online)
    Spatial slices (a) and qualitative surface plots (b) of the CAL trap at various detuning frequencies. In contrast to the bubble trap, the CAL trap is anisotropic, hence the images are shown along different spatial axes. Each consecutive row of images corresponds to a higher detuning frequency, with the first row corresponding to the initial or ``bare" trap.
  }
  \label{CAL_trap}
\end{figure*}

Having found the change in temperature and condensate transition temperature of a system undergoing adiabatic expansion from a filled sphere to a thin shell in an idealized bubble trap, we now consider the expansion of a system in a realistic CAL trap: $V(\vec x)=V_\text{CAL}(\vec x)$.
In the microgravity environment aboard the ISS, CAL is able to produce exceptionally large mm-scale bubbles in both filled and hollow regimes.
Figure.~\ref{CAL_trap} shows a small sample of the CAL dressed potentials experienced by magnetically trapped $^{87}\text{Rb}$ atoms.
As discussed in \cite{Lundblad_2019}, these remarkable dressed potentials are generated using atom-chip current configurations.
By increasing the rf detuning signal, one can deform the trap geometry to produce shell-shaped potential-minimum surfaces.
The CAL trap is thus broadly similar to the bubble trap.
As a result, much of the bubble trap physics discussed above should be applicable to the CAL experiment.

There are crucial differences in the details though; unlike the bubble trap, the CAL trap is spatially inhomogeneous, breaking rotational symmetry. This means quantum-mechanical modelling of CAL trap thermodynamics is significantly more difficult, requiring solutions of three-dimensional partial differential equations as opposed to one-dimensional radial equations.
Fortunately, some of the experimentally-relevant adiabatic expansions occur at temperatures above $T_c$ and hence numerically-expensive diagonalization can be avoided through use of the semiclassical approximation.
We also note that these experimental expansions start with traps at negative detunings which were not considered earlier with the bubble trap. This does not pose a serious problem when comparing condensate physics between the CAL and bubble traps as negative detunings are not differentially meaningful from zero detuning in the sense that condensates in traps below resonance up to resonance form filled geometric structures.

\begin{figure}[htp!]
  \includegraphics[width=\columnwidth]{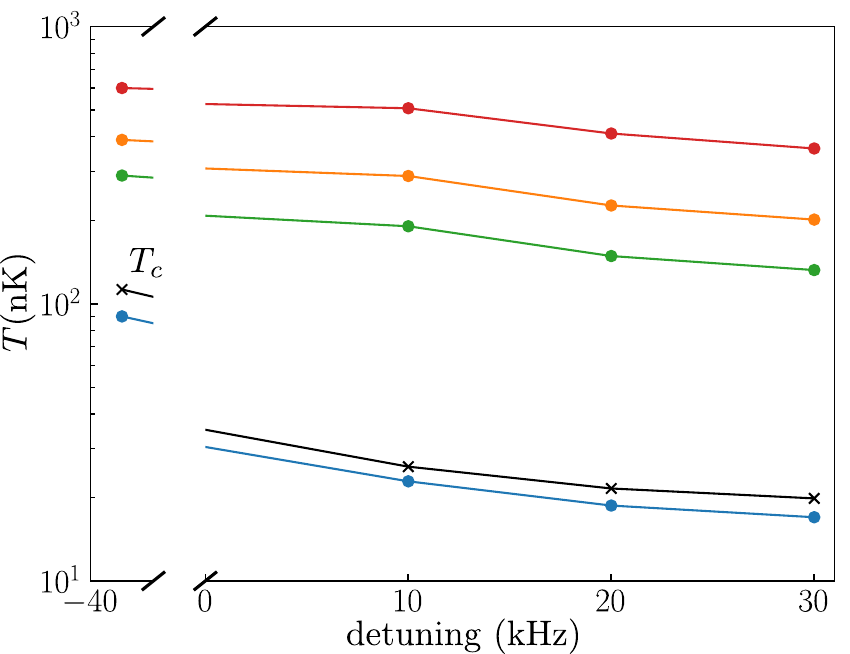}
  \caption{
    (Color online)
    BEC critical temperature (black line) of $N=5\times10^4$ noninteracting $^{87}\text{Rb}$ atoms in the CAL trap along with various adiabatic expansion temperature profiles (colored lines) calculated semiclassically.
    The adiabatic expansions use experimentally-relevant initial temperatures in \cite{Lundblad_2021}: $Ti = 90, 290, 390, 600$ nK.
  }
  \label{CAL_cooling_semiclassical}
\end{figure}

Here we use the semiclassical approach to obtain thermodynamics during system expansion in the CAL trap.
To carry out our calculations, CAL atom-chip current data is used as input to generate a spatial grid of dressed potentials, $V_\text{CAL}(\vec x)$, appropriate for $^{87}\text{Rb}$ atoms in the $|F=2,m_F=2\rangle$ hyperfine state \cite{Lundblad_2019}.
Using the semiclassical approximation for NI atoms, we compute both the BEC critical temperature and the temperature of the system during adiabatic expansions
for initial temperatures relevant to the CAL run reported in \cite{Lundblad_2021} ($Ti = 90, 290, 390, 600$ nK) with the results displayed in Fig.~\ref{CAL_cooling_semiclassical}.
We see both the BEC critical temperature and the temperature during adiabatic expansions decrease with applied detuning frequency. However, $T_c$ does not decrease noticeably faster than the adiabatic curve for the case of the initially-condensed gas. This same phenomenon can be observed in the semiclassical treatment of the bubble trap (see Fig.~\ref{semiclassical_comparison}).
In the case of the bubble trap, we saw that for temperatures at or below $T_c$, the semiclassical approach became less accurate as detuning increased and the bubble became thin.
Most notably, we saw that the semiclassical model overestimates the BEC critical temperature for large bubbles.
Hence, we find it is likely that, at these experimentally-relevant starting temperatures, that adiabatic expansions in the traps on CAL will produce thermal clouds rather than condensates.

\section{Outlook}

In summary, we have presented a thorough study of the thermodynamic properties of BECs in shell-shaped geometries while working in parameter regimes applicable to CAL experiments.
We began by charting out the spectrum and eigenstates of a bubble trap potential through the evolution of trap parameters for a condensate from a filled sphere to hollow thin shell. Based on this analysis, we numerically computed thermodynamic quantities for a noninteracting system and found that adiabatic expansion leads to condensate depletion.
Even in the presence of interactions, we argued that our conclusions hold for fixed particle systems in the thin-shell limit.
Finally, using semiclassical methods to calculate thermodynamic properties of $^{87}\text{Rb}$ atoms in CAL experimental traps, we made the connection between this realistic system and the bubble trap more concrete.

While several immediate avenues open up with regards to BECs in shell-shaped geometries, we first note: the CAL experiments, corroborated by theory, show something remarkable. In the microgravity environment, the shell inflation mechanism provided by the bubble trap allows for macroscopically large, suspended and contained gas structures to emerge.
While macroscopic quantum coherence across these structures is likely to take further efforts, topologically non-trivial thermal gases that span linear dimensions of the order of millimeters could present an alternative source of nK-scale large clouds for atom interferometry, typically obtained otherwise via delta-kick cooling \cite{Becker_2018} or adiabatic expansion \cite{Pollard_2020}.
Our studies show here that it is not impossible to retain condensation during adiabatic expansion into these remarkable bubbles.
The next-generation CAL experiments intend to achieve such condensate shells \cite{Lundblad_2021}, providing fertile ground for future studies.

With regards to equilibrium and linear response features, previous works have elucidated several prospects. These thermodynamic analyses would require further developments to connect with experiments, such as more sophisticated techniques for handling interactions, higher computational power for handling realistic trap geometries, and accounting for the strong asymmetry present in actual traps. Thermodynamic studies would also serve well to interface with other aspects of condensate shells that are being studied, including collective modes and vortex physics. Connecting the thermodynamic studies presented here with other settings for shell-shaped condensates, such as in stellar bodies, optical lattices, and Bose-Fermi mixtures, would offer new insights.

An additional realm poised for investigation involves non-equilibrium dynamics. In this work, we assumed adiabaticity and that the dynamic shell evolution involved a quasi-static process. Generally, there can be a variety of dynamical situations where adiabaticity breaks down; tracking the rate of entropy growth would be one step towards quantifying such deviations.
Perhaps the most drastic deviation from adiabaticity concerns regimes in which the system is dynamically tuned across the BEC transition. Our results do indeed suggest that such transitions during expansion are likely.
In this situation, given that the intrinsic relaxational timescale of the system diverges at criticality, no matter how slow the tuning, adiabaticity breaks down and the system falls out of equilibrium. Such non-equilibrium critical behavior results in the universal Kibble-Zurek scaling of various quantities, such as defect densities, wherein the exponent only depends on critical exponents and the dimensionality of the system \cite{Kibble_1976,Zurek_1985,Dziarmaga_2010,Polkovnikov_2011,Zurek_2014}. In the context of BECs, the non-equilibrium production and scaling of defects, specifically vortices \cite{Zurek_1985,Weiler_2008,DeMarco_2011,Beugnon_2017} and more complex defects in spinful condensates \cite{Sadler_2006,Lamacraft_2007,Saito_2007,Damski_2008,Saito_2013}, has been studied. In the shell situation considered in this work, several variations would come into play. In the expansion considered here, the sphere evolves from a condensed to a thermal gas.
In thereby tuning through the critical point, we do indeed expect to enter a non-adiabatic regime and fall out of equilibrium due to the divergence of the system's intrinsic relaxational timescale \cite{Dziarmaga_2010}, though, we do not expect to see vortices since the system enters an uncondensed phase.
However, reversing the process across the critical point into the condensed phase would show such non-equilibrium production of vortices.
In such tuning through the critical regime, finite size effects would be significant.
Moreover, with the thinning down of the shell, we expect dimensional crossover to occur, providing a new aspect for critical scaling. While these variations are interesting in their own right, the ubiquitous feature that we expect is the breakdown of adiabaticity in tuning across the critical point.

These form but a few considerations in the fascinating and diverse realms that host these unique shell-shaped condensate structures from stellar bodies to systems of co-existent phases on Earth to the on-going studies aboard the International Space Station. A common underlying thread is the extreme conditions and dynamic tuning offered by Nature at the astronomical realms and by advances in the ultracold experiments. One might speculate on how much of the thermodynamics of expanding shells discussed here would be relevant in stellar evolution and formation of neutron stars. In the meanwhile, while we have demonstrated that retaining a condensate through expansion can be a delicate matter, we have shown consistent with hints from CAL, that in the ultracold experimental realm, it is entirely possible to create remarkable, gigantic, diaphanous thermal bubbles.

\begin{acknowledgments}

We are grateful to Karmela Padavi{\'c} and Kuei Sun for involved conversations and laying part of the background work in previous collaborations.
We thank Bryan Clark for illuminating conversations that informed numerics for realistic CAL settings.
This work made use of the Illinois Campus Cluster supported by the University of Illinois at Urbana-Champaign.
This work was supported by the National Aeronautics and Space Administration through a contract with the Jet Propulsion Laboratory, California Institute of Technology.

\end{acknowledgments}

\appendix

\section{Thermodynamics of dilute bosons}\label{thermo_appendix}

Consider the Hamiltonian $\hat{H}$ for a collection of dilute interacting bosons in $\mathbb R^3$, Eq.~(\ref{H_contact_interaction}). We are interested in thermodynamics, hence compute thermal averages using $\langle \cdots \rangle = \text{Tr}(\frac{1}{\mathcal Z} e^{-\beta(\hat H - \mu \hat N)} \cdots)$,  where the trace is over Fock space, $\beta(=1/k_B T)$ is the inverse temperature, $\mu$ is the chemical potential, $\hat N$ is the number operator, and $\mathcal Z$ is the grand partition function:
\begin{eqnarray}
    \mathcal Z = \text{Tr} \left( e^{-\beta(\hat H - \mu \hat N)} \right) = \mathcal N \int [D\psi^* D\psi] \, e^{- S[\psi^* \! , \psi]/\hbar}
    \label{grand_partition_fn}.\nonumber\\
    \,
\end{eqnarray}
Above, we introduced the coherent state path integral in the usual way \cite{Altland_2010}: $\mathcal N$ is an unimportant constant prefactor, $[D\psi^* D\psi]$ is the integration measure of a complex-valued field $\psi(\tau, \vec x)$ with ``time" coordinate $\tau\in [0,\beta\hbar)$ (the field obeys periodic boundary conditions in time), and the action is
\begin{eqnarray}
    S = \int d\tau d^3 x \left[ \psi^* \left( \hbar \partial_\tau-\frac{\hbar^2}{2m}\nabla^2 + V -\mu \right) \psi + \frac{g}{2} |\psi|^4 \right]
    \label{S_contact_interaction}.
    \nonumber\\
    \,
\end{eqnarray}
As $T\to0^+$ ($\beta\to\infty$), only field configurations which minimize the action contribute significant weight to the partition function. We denote such configurations as ``condensate wavefunctions" $\psi_c$ \cite{Pethick_2008,Pitaevskii_2016} --- they are static and obey the time-independent Gross-Pitaevskii equation, Eq.~(\ref{GP}).

\subsection{Thermodynamics of noninteracting bosons}

With interactions turned off ($g=0$), thermodynamics are obtained from the spectrum and wavefunctions of the single-particle Schr{\"o}dinger equation \cite{Pethick_2008,Pitaevskii_2016}. In this section, let ``$\alpha$" ($\varepsilon_\alpha$) denote the eigenstates (eigenvalues) of the Schr{\"o}dinger equation:
\begin{eqnarray}
    \left( -\frac{\hbar^2}{2m}\nabla^2 + V(\vec x) \right) \phi_\alpha(\vec x)=\varepsilon_\alpha \phi_\alpha(\vec x)
    \label{single_particle_Schrodinger}.
\end{eqnarray}
Further, let $\alpha=0$ denote the ground state of this equation (which we assume is non-degenerate) and $\varepsilon_0$ denote the ground state energy. Expanding the field operator as $\hat \psi (\vec x) = \sum_\alpha \phi_\alpha(\vec x) \hat b_\alpha$ in the operator formalism or the field as $\psi(\tau,\vec x) = \sum_{n \alpha} e^{-i \omega_n \tau} \phi_\alpha(\vec x) \psi_{n\alpha}$, where $\omega_n$ are bosonic Matsubara frequencies \cite{Altland_2010}, in the path integral formalism gives
\begin{eqnarray}
    \mathcal Z = \prod_\alpha \frac{1}{1-e^{-\beta(\varepsilon_\alpha-\mu)}}
    \label{NI_partition_fn},
\end{eqnarray}
where $\mu<\varepsilon_0$ in order to keep the trace bounded. From derivatives of the grand free energy, $ - \frac{1}{\beta} \ln \mathcal Z$, various thermal expectation values can be computed:
\begin{subequations}
    \label{NI_thermodynamics_appendix}
    \begin{eqnarray}
        N &&= \sum_\alpha f_\alpha
        \label{NI_N_appendix},
        \\
        E &&= \sum_\alpha f_\alpha \varepsilon_\alpha
        \label{NI_E},
        \\
        S &&= k_B \sum_\alpha \left[ (1+f_\alpha)\ln(1+f_\alpha) - f_\alpha \ln f_\alpha \right]
        \label{NI_S_appendix},
    \end{eqnarray}
\end{subequations}
where we denote the system's particle number, energy, and entropy by $N$, $E$, and $S$ respectively, and $f_\alpha=1/(e^{\beta(\varepsilon_\alpha-\mu)}-1)$ is the Bose-Einstein distribution function. If we move onto spatially resolved quantities, such as the one-body density matrix,
\begin{eqnarray}
    \langle \hat\psi^\dagger(\vec x) \hat\psi(\vec x') \rangle = \sum_\alpha \phi_\alpha^*(\vec x) \phi_\alpha(\vec x') f_\alpha
    \label{NI_1_body_density_matrix},
\end{eqnarray}
we see these require information on the wavefunctions. Thus, at least formally, solving Eq.~(\ref{single_particle_Schrodinger}) gets us the thermodynamics.

\subsection{The semiclassical approximation}

When solutions to Eq.~(\ref{single_particle_Schrodinger}) are intractable a useful approximation exists. Provided the temperature is much larger than the single-particle energy level spacing, one can approximate thermodynamic quantities by replacing sums over eigenstates with integrals: $\sum_\alpha\to``\int d\alpha"$ \cite{Pethick_2008,Pitaevskii_2016}.
Unfortunately, one can only make this replacement if they have a notion of what the states in Eq.~(\ref{single_particle_Schrodinger}) actually are.
However, at temperatures large enough that the phase-space density is small, one can instead work with the classical relation $\varepsilon_{\vec p} (\vec x) = \frac{1}{2m} |\vec p|^2 + V(\vec x)$ \cite{Pethick_2008,Pitaevskii_2016}.
Provided the system is large enough to ignore boundary effects and treat momentum as continuous, the semiclassical approximation is to make the following replacement:
\begin{eqnarray}
    \sum_\alpha F(\varepsilon_\alpha) \to \int \frac{d^3x d^3p}{(2\pi\hbar)^3} F(\varepsilon_{\vec p} (\vec x))
    \label{semiclassical_approximation},
\end{eqnarray}
where $F(\cdot)$ is a function of the dispersion. In the normal phase, $T>T_c$, one finds (after integrating over momentum) the semiclassical expressions of the thermodynamic sums in Eq.~(\ref{NI_thermodynamics_appendix}) are:
\begin{subequations}
    \label{SC_thermodynamics}
    \begin{eqnarray}
        N &&=
        n_Q \int_{\mathbb{R}^3} d^3 x \, \text{Li}_{\frac{3}{2}}(e^{-\beta(V-\mu)})
        \label{SC_N},
        \\
        E &&=
        n_Q \int_{\mathbb{R}^3} d^3 x
        \bigg[
        \, \frac{3}{2} k_B T \, \text{Li}_{\frac{5}{2}}(e^{-\beta(V-\mu)})
        \nonumber\\
        &&\qquad\qquad\qquad\quad + V \, \text{Li}_{\frac{3}{2}}(e^{-\beta(V-\mu)})
        \bigg]
        \label{SC_E},
        \\
        S &&= k_B
        n_Q \int_{\mathbb{R}^3} d^3 x
        \bigg[
        \, \frac{5}{2} \, \text{Li}_{\frac{5}{2}}(e^{-\beta(V-\mu)})
        \nonumber\\
        &&\qquad\qquad\qquad\quad + \beta(V-\mu) \, \text{Li}_{\frac{3}{2}}(e^{-\beta(V-\mu)})
        \bigg]
        \label{SC_S},\qquad
    \end{eqnarray}
\end{subequations}
where $n_Q(T) \equiv \lambda_T^{-3} = (m k_B T / 2\pi \hbar^2)^{3/2}$ is the quantum density and $\text{Li}_{s}(z) = \sum_{n=1}^\infty z^n / n^{s}$ is the polylogarithm or, as commonly referred to in this context, the Bose function \cite{Houbiers_1997}.

\subsection{Thermodynamics of (weakly) interacting bosons at low-temperatures}

With interactions turned on ($g>0$), computing the thermodynamics is considerably more difficult \cite{Andersen_2004,Yukalov_2004}.
In general, path integral Monte Carlo methods offer a powerful way to proceed \cite{Ceperley_1995,Ceperley_1997}.
Green function methods have found success in addressing questions such as the shift in $T_c$ from its noninteracting value \cite{Baym_1999,Baym_2000,Holzmann_2001}.
At temperatures well below (or commensurate to) the BEC critical temperature, it is common to employ some variation of the mean-field theory introduced by Bogoliubov \cite{Bogoliubov_1947,Pethick_2008,Pitaevskii_2016,Yukalov_2014,Tononi_2019}. Because our interests lie in ultracold dilute atomic gases, we will restrict ourselves to temperatures far below the BEC transition, $T \ll T_c$, and use a Bogoliubov quasiparticle description.

At sufficiently low temperatures, provided the dimension of space is larger than the lower critical dimension (which is $2$ from the Mermin-Wagner–Hohenberg \cite{Mermin_1966,Hohenberg_1967} or Coleman \cite{Coleman_1973} theorem), the system can spontaneously break the $U(1)$ symmetry of the Hamiltonian. This is signified by a nonvanishing field expectation value: $\langle \hat\psi \rangle \sim \psi_c$ as $T\to 0$ where the ground state field configuration, $\psi_c$, is a solution to Eq.~(\ref{GP}).
At these low-temperatures, we perform a saddle-point analysis to reach an effective low-temperature theory for fluctuations \textit{out of the condensate}.
Namely, we use the ``Bogoliubov shift": $\psi = \psi_c + \delta\psi$ \cite{Yukalov_2014,Tononi_2019}. In the operator formalism:
\begin{eqnarray}
    \hat H - \mu \hat N &&= -E^c_\text{int}
    \nonumber\\
    &&+ \int_{\mathbb R^3} d^3 x \,
    \delta\hat\psi^\dagger \left( -\frac{\hbar^2}{2m}\nabla^2 + V - \mu + 2g |\psi_c|^2 \right) \delta\hat\psi
    \nonumber\\
    &&+ \int_{\mathbb R^3} d^3 x \,
    \frac{g}{2} \left( \psi_c^2 \delta\hat\psi^\dagger \delta\hat\psi^\dagger + \text{h.c.}  \right)
    + \mathcal{O} ( \delta\hat\psi )^3
    \label{H_expansion},
\end{eqnarray}
where $E^c_\text{int} \equiv \frac{g}{2} \int d^3 x |\psi_c|^4$ is the interaction energy of the condensate wavefunction and the fluctuations, $\delta\hat\psi$, obey bosonic commutation relations. The quadratic portion of the operator in Eq.~(\ref{H_expansion}) can be brought into diagonal form by use of a Bogoliubov transformation \cite{Pethick_2008,Pitaevskii_2016,Blaizot_1986}:
\begin{eqnarray}
    \hat H - && \mu \hat N= -E^c_\text{int}
    \nonumber\\
    &&
    +
    E_\text{pair}
    +
    \frac{1}{2M}\hat{P}_z^2
    +
    \sum_{\substack{\lambda \\ (E_\lambda > 0)}} E_\lambda \hat \beta_\lambda^\dagger \hat \beta_\lambda
    + \mathcal{O} ( \delta\hat\psi )^3
    \label{H_bogoliubov}.\qquad
\end{eqnarray}
In the equation above, the energies, $E_\lambda$, of the (bosonic) quasiparticles, $\hat \beta_\lambda$, are given by solving the Bogoliubov equations \cite{Pethick_2008,Pitaevskii_2016}:
\begin{subequations}
    \label{Bogo_eqs}
    \begin{eqnarray}
        E_\lambda u_\lambda(\vec x) &&= \left( -\frac{\hbar^2}{2m}\nabla^2 + V - \mu + 2g |\psi_c|^2 \right) u_\lambda(\vec x)
        \nonumber\\
        &&+ g \psi_c^2 v_\lambda(\vec x)
        \label{Bogo_1},
        \\
        -E_\lambda v_\lambda(\vec x) &&= \left( -\frac{\hbar^2}{2m}\nabla^2 + V - \mu + 2g |\psi_c|^2 \right) v_\lambda(\vec x)
        \nonumber\\
        &&+ g \psi_c^{*2} u_\lambda(\vec x)
        \label{Bogo_2},
    \end{eqnarray}
\end{subequations}
where we introduced the wavefunctions $u_\lambda(\vec x)$ and $v_\lambda(\vec x)$. For positive energy solutions, the wavefunctions can be chosen to obey the orthonormality conditions \cite{Blaizot_1986}:
\begin{subequations}
    \label{orthonormal}
    \begin{eqnarray}
        \delta_{\lambda \lambda'} =
        \int_{\mathbb R^3} d^3 x \left(
        u_{\lambda}^* u_{\lambda'} - v_{\lambda}^* v_{\lambda'} 
        \right)
        \label{orthonormal_1},
        \\
        0 =
        \int_{\mathbb R^3} d^3 x \left(
        u_{\lambda} v_{\lambda'} - v_{\lambda} u_{\lambda'}
        \right)
        \label{orthonormal_2},
    \end{eqnarray}
\end{subequations}
The Bogoliubov collective excitations in Eq.~(\ref{H_bogoliubov}) are created by:
\begin{eqnarray}
    \hat \beta_\lambda^\dagger \equiv \int_{\mathbb{R}^3} d^3x \left( u_\lambda \delta\hat\psi^\dagger - v_\lambda \delta\hat\psi \right)
    \label{quasiparticle},
\end{eqnarray}
and thus $u_\lambda(\vec x)$ and $v_\lambda(\vec x)$ have the interpretation of a ``particle" and ``hole" wavefunction respectively. We also see the emergence of a ``momentum operator" \cite{Blaizot_1986,Pethick_2008},
\begin{eqnarray}
    \hat P_z \equiv \int_{\mathbb{R}^3} d^3x \left( \psi_c \, \delta\hat\psi^\dagger + \text{h.c.} \right)
    \label{P_z},
\end{eqnarray}
associated with a zero energy eigenvalue (or zero-mode) solution of the Bogoliubov equations: $u_z(\vec x)=-v_z^*(\vec x)=\psi_c(\vec x)$ \cite{Blaizot_1986,Lewenstein_1996}.
The ``mass" in the Hamiltonian, $M \equiv \partial_\mu \int d^3 x |\psi_c|^2$, is a (positive) inverse energy scale introduced for convenience \cite{Blaizot_1986}. Lastly, in Eq.~(\ref{H_bogoliubov}) there is a shift in the reference energy that arises due to the pairing terms in Eq.~(\ref{H_expansion}). This energy shift, which is nonpositive, is given by:
\begin{eqnarray}
    E_\text{pair} \equiv - \int_{\mathbb{R}^3} d^3x \bigg(
    \frac{1}{2M} |\psi_c|^2
    +
    \sum_{\substack{\lambda \\ (E_\lambda > 0)}} E_\lambda |v_\lambda|^2
    \bigg)
    \label{E_pair}.\qquad
\end{eqnarray}

We can define a ``position operator", $\hat Q_z$, which obeys the canonical commutation relation $[\hat Q_z, \hat P_z] = i$ as \cite{Lewenstein_1996,Pethick_2008}
\begin{eqnarray}
    \hat Q_z \equiv \int_{\mathbb{R}^3} d^3x \bigg( -\frac{i \partial_\mu\psi_c }{M} \, \delta\hat\psi^\dagger + \text{h.c.} \bigg)
    \label{Q_z}.
\end{eqnarray}
Note: the zero-mode operators $\hat P_z$ and $\hat Q_z$ commute with the $\{\hat \beta_\lambda\}$ in Eq.~(\ref{H_bogoliubov}). From the new operators, the original fluctuation field operator can be written as \cite{Lewenstein_1996}
\begin{eqnarray}
    \delta\hat\psi
    = \frac{\partial_\mu \psi_c}{M} \, \hat P_z - i \psi_c \, \hat Q_z
    +
    \sum_{\substack{\lambda \\ (E_\lambda > 0)}} (u_\lambda \hat \beta_\lambda + v_\lambda^* \hat \beta_\lambda^\dagger)
    \label{fluctuation}.\qquad
\end{eqnarray}

For temperatures far below the $U(1)$ symmetry breaking transition, we assume that, when calculating observables, it is permissible to ignore the third and fourth order fluctuation terms in Eq.~(\ref{H_bogoliubov}). For consistency, this requires that the ``condensate depletion" is small: $\langle \int d^3 x \delta\hat\psi^\dagger \delta\hat\psi \rangle / N \ll 1$. Thus, we reach an effective Hamiltonian for the fluctuations:
\begin{eqnarray}
    \hat H_\text{eff} \equiv
    -E^c_\text{int}
    +E_\text{pair}
    +
    \frac{1}{2M}\hat{P}_z^2
    +
    \sum_{\substack{\lambda \\ (E_\lambda > 0)}} E_\lambda \hat \beta_\lambda^\dagger \hat \beta_\lambda
    \label{H_eff}.\qquad
\end{eqnarray}

The eigenstates of Eq.~(\ref{H_eff}) are simply those of the $\hat P_z$ operator and quasiparticle number operators:
\begin{eqnarray}
    |P_z , \{n_\lambda\}\rangle
    =
    \frac{  e^{-\frac{1}{2}P_z^2 + i \sqrt{2} P_z \hat \beta_z^\dagger + \frac{1}{2}(\hat \beta_z^\dagger)^2}  }{\pi^{1/4}}
    \prod_{\substack{\lambda \\ (E_\lambda > 0)}} \frac{(\hat \beta_\lambda^\dagger)^{n_\lambda}}{\sqrt{n_\lambda !}}
    |\text{vac}\rangle
    \label{H_eff_eigenstates},
    \nonumber\\
    \,
\end{eqnarray}
where $P_z\in\mathbb{R}$, $n_\lambda\in\mathbb{Z}_{\ge 0}$, $\hat \beta_z$ is a ladder operator associated with the zero-mode operators \cite{Lewenstein_1996} defined as $\hat \beta_z \equiv (\hat Q_z + i \hat P_z)/\sqrt{2}$, and $|\text{vac}\rangle$ is the vacuum state of the $\hat \beta_z$ and $\{\hat \beta_\lambda\}$ operators.

We should now be in a position to approximate thermal averages in the broken $U(1)$ phase using $\langle \cdots \rangle \approx \text{Tr}(\frac{1}{\mathcal Z} e^{-\beta \hat H_\text{eff}} \cdots)$, but as discussed in \cite{Lewenstein_1996} doing so naively leads to inconsistent results. Because the eigenstates of the effective Hamiltonian have \textit{definite} zero-mode ``momentum", i.e. $P_z$ is a good quantum number, we necessarily find $\langle \hat Q_z^2\rangle$ diverges as a result of the Heisenberg uncertainty principle. Because $\langle \delta\hat\psi^\dagger \delta\hat\psi \rangle$ contains a term proportional the variance in the $\hat Q_z$ operator, we then violate the condition that the condensate depletion be small \cite{Lewenstein_1996}.
Thus, when using the effective Hamiltonian to compute thermal expectation values, one cannot simultaneously trace over the zero-mode ``momentum" states and demand the depletion be small.

Here, we avoid this issue in a pragmatic way through our choice of the condensate wavefunction.
We consider cases where the condensate wavefunction is well approximated by the noninteracting (NI), $g=0$, solution: $\psi_c(\vec x) \approx \psi_\text{NI}(\vec x) = \sqrt{N_0} \phi_0 (\vec x)$.
Although this may seem like a poor choice, we find in the main text that using the NI condensate wavefunction can be justified in the limit of thin-shells at fixed particle number.
If one replaces $\psi_c(\vec x)$ with $\sqrt{N_0} \phi_0 (\vec x)$ one finds both $\hat P_z$ and $\hat Q_z$ vanish if one further neglects fluctuations in the $\alpha=0$ mode: $\delta \hat b_0 \equiv \hat b_0 - \sqrt{N_0} \to 0$.
Treating $\hat b_0$ as a c-number is well-justified for $T\ll T_c$ where $N_0$ is macroscopically large; thus, the approximation is standard in the Bogoliubov treatment of the weakly interacting Bose gas \cite{Bogoliubov_1947,Pethick_2008,Pitaevskii_2016}.
To summarize, in the $U(1)$ symmetry broken phase, if we assume 1) the gas is sufficiently dilute that the NI condensate wavefunction is a reasonable approximation of the true condensate wavefunction and 2) the temperature is far enough below the BEC transition that it is acceptable to ignore fluctuations in the NI ground state mode, our effective Hamiltonian is still given by Eq.~(\ref{H_eff}), but with the $\hat P_z$ term simply removed.

Under these assumptions, we can now calculate thermodynamic quantities in the presence of (weak) interactions. First, the field obtains a nonzero expectation value, $\langle \hat\psi (\vec x) \rangle = \psi_c(\vec x)$, spontaneously breaking the $U(1)$ symmetry of the Hamiltonian.
This also states the condensate wavefunction is the mean-field in the broken symmetry phase for all temperatures at which the effective Hamiltonian remains valid. Next, the one-body density matrix becomes
\begin{eqnarray}
    \langle &&  \hat\psi^\dagger(\vec x) \hat\psi(\vec x') \rangle
    =
    \psi_c^*(\vec x) \psi_c(\vec x')
    \nonumber\\
    &&+
    \sum_{\substack{\lambda \\ (E_\lambda > 0)}}
    \left[ f_\lambda u_\lambda^*(\vec x) u_\lambda(\vec x ') + (1 + f_\lambda) v_\lambda(\vec x) v_\lambda^*(\vec x ') \right]
    \label{1_body_density_matrix},\qquad
\end{eqnarray}
where $f_\lambda=1/(e^{\beta E_\lambda}-1)$ is the Bose-Einstein distribution function for the Bogoliubov quasiparticles.
It is then straightforward to show in the thermodynamic limit that the number of particles is:
\begin{eqnarray}
    && N = N_0(T)
    \nonumber\\
    && +
    \sum_{\substack{\lambda \\ (E_\lambda > 0)}}
    \int_{\mathbb R^3} d^3 x
    \left[ f_\lambda |u_{\lambda}|^2 + (1+f_\lambda) |v_{\lambda}|^2 \right]
    \label{bogo_N}.
\end{eqnarray}
Notice that to ensure self-consistency at temperature $T$, the number of particles in the NI ground state, $N_0(T)$, must be chosen such that Eq.~(\ref{bogo_N}) is satisfied.

\section{Bogoliubov equations in the bubble trap}\label{BOGO_bubble_appendix}

Within the mean-field theory discussed in appendix \ref{thermo_appendix}, thermodynamics of the interacting system follow from solutions to the Bogoliubov equations, Eq.~(\ref{Bogo_eqs}).
Due to rotational invariance of the bubble trap, $V(r)$, and the condensate wavefunction, $\psi_c(r)$, we can write Bogoliubov ``particle" and ``hole" wavefunctions respectively as
\begin{subequations}
    \label{Bogo_wavefns_radial}
    \begin{eqnarray}
        u_{kl{m_l}}(\vec x) && =\frac{1}{r}u_{k l}(r) Y_l^{m_l}(\theta,\phi)
        \label{Bogo_u_radial},
        \\
        v_{kl{m_l}}(\vec x) && =\frac{1}{r}v_{k l}(r) Y_l^{m_l}(\theta,\phi)
        \label{Bogo_v_radial},
    \end{eqnarray}
\end{subequations}
where the notation is the same as that used in Eq.~(\ref{single_particle_wavefns}). Here the radial wavefunctions $u_{kl}(r)$ and $v_{kl}(r)$ (which obey open boundary conditions) satisfy the Bogoliubov analog of the radial equation:
\begin{subequations}
    \label{Bogo_radial_eqs}
    \begin{eqnarray}
        E_{kl} u_{kl} &&=
        \left( -\frac{\hbar^2}{2m}\frac{d^2}{dr^2} + V + \frac{\hbar^2 l(l+1)}{2m r^2} - \mu + 2g|\psi_c|^2 \right) u_{k l}
        \nonumber\\
        &&+ g \psi_c^2 \, v_{kl}
        \label{Bogo_1_radial},
        \\
        -E_{kl} v_{kl} &&=
        \left( -\frac{\hbar^2}{2m}\frac{d^2}{dr^2} + V + \frac{\hbar^2 l(l+1)}{2m r^2} - \mu + 2g|\psi_c|^2 \right) v_{k l}
        \nonumber\\
        &&+ g {\psi_c^*}^2 \, u_{kl}
        \label{Bogo_2_radial}.
    \end{eqnarray}
\end{subequations}
The number of particles in the $U(1)$ symmetry broken phase is then
\begin{eqnarray}
    && N = N_0(T)
    \nonumber\\
    &&
    +
    \sum_{\substack{kl \\ (E_{kl} > 0)}} (2l + 1) \int_0^\infty dr
    \bigg[ \, 
    f_{kl} |u_{kl}|^2
    + 
    (1+f_{kl}) |v_{kl}|^2
    \bigg]
    \label{bogo_N_radial},\qquad
\end{eqnarray}
where $f_{kl} = 1/(e^{\beta E_{kl}}-1)$.

To solve Eq.~(\ref{Bogo_radial_eqs}), one must set the condensate wavefunction and chemical potential. In principle, this is done by first solving the time-independent Gross-Pitaevskii (GP) equation, Eq.~(\ref{GP}); however, within our mean-field theory we treat the noninteracting (NI) solution of the GP equation as a reasonable approximation. If one takes $g\to 0$ in the GP equation, the condensate wavefunction simplifies to $\psi_c(r) \to \psi_\text{NI}(r) = \sqrt{N_0} \, u^\text{NI}_0(r) /  \sqrt{4\pi} \, r $, where $u^\text{NI}_0(r)$ is the ground state solution of the radial equation, Eq.~(\ref{radial_eq}), and the chemical potential takes on the form $\mu\to\varepsilon_0$.
Returning to the interacting case, we insert the approximate condensate wavefunction $\psi_\text{NI}(r)$ into the Bogoliubov equations and set the chemical potential $\mu$ such that the $k=0$, $l=0$, ${m_l}=0$ mode of Eq.~(\ref{Bogo_radial_eqs}) is a zero-mode. This is done because the true condensate wavefunction $\psi_c(r)$ corresponds to a zero-mode solution of the Bogoliubov equations.

\bibliography{bibliography}

\end{document}